# Origin of interfacial conductivity at complex oxide heterointerfaces: possibility of electron transfer from water chemistry at surface oxygen vacancies

Meng Zhang[1,2], Zheng Chen[1,2], Baohua Mao[3], Qingtian Li[3], Hai Bo[1,2], Tianshuang Ren[1,2], Pimo He[1,2], Zhi Liu,[3,4] and Yanwu Xie[1,2*]

[1]Department of Physics, Zhejiang University, Hangzhou 310027, China

[2]Collaborative Innovation Center of Advanced Microstructures, Nanjing University, Nanjing 210093, China

[3]State Key Laboratory of Functional Materials for Informatics, Shanghai Institute of Microsystem and Information Technology, Chinese Academy of Sciences, Shanghai 200050, China

[4]School of Physical Science and Technology, ShanghaiTech University, Shanghai 200031, China

[*]Correspondence and requests for materials should be addressed to Y. X. (email: ywxie@zju.edu.cn)

**Abstract**

Variety of conducting heterointerfaces have been made between $SrTiO_3$ substrates and thin capping layers of distinctly different oxide materials that can be classified into polar band insulators (e.g. $LaAlO_3$), polar Mott insulators (e.g. $LaTiO_3$), apparently non-polar band insulators (e.g. $\gamma$-$Al_2O_3$), and amorphous oxides (e.g. amorphous $SrTiO_3$). A fundamental question to ask is if there is a common mechanism that governs interfacial conductivity in all these heterointerfaces. Here, we examined the conductivity of different kinds of heterointerfaces by annealing in oxygen and surface treatment with water. It was found that the conductivity of all the heterointerfaces show a strong dependence on annealing, and can be universally tuned by surface treatment whose effect is determined by the annealing condition. These observations, together with ambient-pressure X-ray photoelectron spectroscopy measurements, suggest that water chemistry at surface oxygen vacancies is a common mechanism that supplies electrons to the interface.

## I. INTRODUCTION

The conducting heterointerfaces formed between insulating complex oxides have attracted intense research interest in both the fundamental physics of interfacial emergent phenomena [1] and the potential applications in oxide electronics [2,3]. A typical heterointerface is generated by growing a thin (a few nanometers thick) capping oxide layer on a $SrTiO_3$ single crystal substrate. The conductance locates at the $SrTiO_3$ side near the interface, and the capping layer works as a medium to induce interfacial electrons. The best known example is the $LaAlO_3/SrTiO_3$ heterointerface reported on 2004 [4], in which a polar $LaAlO_3$ layer was deposited on a non-polar $TiO_2$ terminated (100) $SrTiO_3$ substrate (the most widely used oxide substrate). A heuristic polar catastrophe and electronic reconstruction mechanism was widely used to explain the origin of the $LaAlO_3/SrTiO_3$ interfacial conductivity [5], while oxygen vacancies in $SrTiO_3$ have also been realized to play an important role [6]. Theoretically, a polarity-induced defect mechanism [7] and surface hydrogen adsorption [8] or redox reaction [9] have also been proposed to explain the interfacial conductivity. Experimentally, it was found that the interfacial conductivity of $LaAlO_3/SrTiO_3$ can be dramatically tuned from the $LaAlO_3$ surface, by charged scanning probe [10–12] or adsorbates [13–15]. A recent X-ray photoelectron spectroscopy study even suggests that surface hydrogen adsorption, provided by the tiny remaining water in $O_2$ gas in the growth chamber, is the dominant source for the conducting electrons [16]. Moreover, it is gradually realized that the non-idealness in real samples, such as interfacial cation mixing and non-stoichiometry in $LaAlO_3$, has a significant influence on the interfacial conductivity [17]. One remarkable observation is that the most high-quality $LaAlO_3/SrTiO_3$ heterostructures with good $LaAlO_3$ stoichiometry are insulating [18] (or less conducting [19]), while similar heterostructures with Al-rich $LaAlO_3$ films are highly conducting, hinting that the non-idealness in real samples is a key factor for the formation of interfacial conductivity. Until now, a shared view is still lacking.

While the $LaAlO_3/SrTiO_3$ heterostructure has attracted overwhelming research focus, a few new capping materials other than $LaAlO_3$ have been used to make similar conducting heterointerfaces [20–29]. These materials vary largely from polar band insulators (e.g. $NdGaO_3$ [21] and $(La_{0.3}Sr_{0.7})(Al_{0.65}Ta_{0.35})O_3$ (LSAT) [24]), polar Mott insulators (e. g. $LaTiO_3$ [26] and $LaVO_3$ [25]), apparently non-polar band insulators (e.g. $\gamma$-$Al_2O_3$ [29] and $CaZrO_3$ [23]), to amorphous oxides [22] (e.g. amorphous $SrTiO_3$ and $LaAlO_3$). In some of these heterointerfaces the polar catastrophe and electronic reconstruction idea as in $LaAlO_3/SrTiO_3$ was used to explain the formation of conducting interfaces [21,23–26], while in others the creation of oxygen vacancies in $SrTiO_3$, e.g. by redox reactions, was proposed [22,29,30]. However, the polar catastrophe and electronic reconstruction mechanism cannot explain the conductivity of heterointerfaces with non-polar capping layers, as well as of $LaAlO_3$ (110)/$SrTiO_3$ that has no polar discontinuity at the interface [31,32]; the oxygen vacancy mechanism alone cannot explain many critical findings of complex oxide heterointerfaces [3,7]. Given the large variety of capping materials and their distinct differences in physical properties, an interesting question arises: Is there a general mechanism that governs, or at least contributes to, the formation of the interfacial conductivity in all these complex oxide heterointerfaces? To address this question, one must find out the common features shared by all of these different heterointerfaces.

In this study, we examined a variety of heterointerfaces that have been reported to be conducting. We investigated their response to the annealing in oxygen and the surface treatment with water (as well as acetone and ethanol). We found that most of the heterointerfaces become insulating after annealing at a moderate temperature whose value varies largely for different capping oxides; surface treatment can universally increase the conductivity in heterointerfaces of almost all kinds of capping oxides except for the amorphous materials, of which, in contrast, the conductivity was decreased; for a given heterointerface, annealing can remarkably increase its sensitivity to surface treatment. The observed close interplay between annealing and surface treatment suggests strongly that the interfacial conductivity is controlled by water-adsorption-related chemistry occurring at the surface oxygen vacancies. The common features shared by all the heterointerfaces are the surface oxygen vacancies and the water chemistry, both of which are almost unavoidable for all the real samples. This new understanding naturally links the electronic reconstruction, oxygen vacancies, and surface chemistry, which have been heavily discussed in previous studies, and can be used to consistently explain most of the present and previous observations, including these that had caused great puzzles, with a simple electrostatic consideration.

## II. EXPERIMENTAL DETAILS

### A. Sample growth.

All samples presented here were prepared by growing capping layers on $SrTiO_3$ single crystal substrates by pulsed laser deposition. To ensure that they are comparable to those reported in previous literatures [4,20–29], we used the most typical growth conductions as reported previously. The $LaAlO_3$(110)/$SrTiO_3$ and $LaAlO_3$(111)/$SrTiO_3$ heterointerfaces were grown on (110) and (111) $SrTiO_3$ substrates, respectively. All others were grown on $TiO_2$-terminated (100) $SrTiO_3$ substrates. Before growth, the (110) and (111) $SrTiO_3$ substrates were treated in a tube furnace at 1100 °C for 2 hours under about 1 bar oxygen; the (100) $SrTiO_3$ substrates were pre-annealed in situ at 975 °C, $10^{-4}$ mbar oxygen, for around 20 minutes. The film thickness was monitored either by *in-situ* reflection high-energy electron diffraction (RHEED), or by counting the pulses of deposition laser. More details are listed in the Table 1. RHEED and atomic force microscopy characterization of typical samples were shown in Figs. S1-S14 in the Supplemental Material(SM) [33]. They are comparable to those reported previously and atomic flat surface were observed [33]. While most of our heterointerfaces show comparable transport properties with those reported in literatures, we note that the sample quality of the $LaAlO_3$ (110)/$SrTiO_3$ and $LaAlO_3$ (111)/$SrTiO_3$ heterointerfaces is somewhat inferior to that reported in the literature in which the metallic conductivity can survive an *in-situ* annealing in oxygen [32].

### B. Post-annealing in oxygen.

The as-grown $^*LaAlO_3$/$SrTiO_3$, $^*LaAlO_3$ (111)/$SrTiO_3$, and $^*$LSAT/ $SrTiO_3$ samples were annealed *in situ* in the growth chamber, as shown in the Table I. All other samples were annealed, when mentioned, *ex situ* in a tube furnace, under a pure oxygen flow slightly above 1 bar, at a temperature of 300-600 °C, for 1 hour.

## C. Electrical measurement.

The electrical contacts to the conducting heterointerfaces were made with Al wires by ultrasonic bonding machine that can penetrate the insulating capping layers. Good Ohmic contacts were achieved due to the small work function of Al. Temperature-dependent transport measurements were performed in a cryostat. Room temperature measurements were performed in ambient environment.

## D. Surface treatment with water.

The processes of surface treatments were done by dropping liquid water on the samples with a pipette firstly, and then drying the samples immediately by blowing them with a nitrogen gas gun (as shown in Fig. S15). All the results shown in this study were obtained by treating and measuring samples at ambient conditions. However, we emphasize that we have performed both the surface treatments and the following measurements in dark and observed essentially the same results, confirming that illumination is not a key issue in the present study [14].

## E. AP-XPS measurement.

Ambient Pressure X-ray Photoelectron Spectroscopy (AP-XPS) [34] measurements were performed in a near ambient pressure facility with laboratory monochromatic Al *K*α x-ray source, with an overall energy resolution of approximately 0.4 eV (a binding energy (BE) shift of ~0.1 eV is detectable). The c-LAO and a-LAO samples were fabricated by pulsed laser deposition *ex situ*, and transferred simultaneously into the XPS chamber through air, without any in-situ cleaning process before measurements (labelled as "as in"). To avoid charging effect during AP-XPS measurement, the metallic interfaces of samples were carefully grounded with Al wires by ultrasonic bonding machine. The base pressure of the XPS chamber is better than $5\times10^{-10}$ mbar. The samples were measured in both base pressure and a water ($H_2O$) vapor of 0.3 mbar. All data were recorded at room temperature. We rule out the possibility that the observed BE shifts are related to any uncompensated charging effect since if they are from charge effect, (1) all core levels should have a similar BE shift (*vs* only core levels in $LaAlO_3$ layer have significant shifts); (2) applying water vapor will reduce the charge effect and thus shift core levels back to lower BE (*vs* no shift or shift to higher BE).

# III RESULTS

## A. Transport properties of the as-grown and annealed heterointerfaces.

Figure 1a shows the temperature dependence of sheet resistivity, $R_{sheet}$, of a variety of as-grown heterointerfaces. Because most of the reported conducting heterointerfaces, except for $LaAlO_3/SrTiO_3$ and $LSAT/SrTiO_3$, were grown without in-situ post-annealing in high oxygen pressure, we prepared all the heterointerfaces without in-situ post-annealing as well. For comparison, in Figure 1a we also include the result of $LaAlO_3/SrTiO_3$ and $LSAT/SrTiO_3$ heterointerfaces with in-situ post-annealing, labeled as $^{*}LaAlO_3/SrTiO_3$ and $^{*}LSAT/SrTiO_3$, since they are the most widely studied ones. All of these heterointerfaces exhibit metallic behavior over a wide temperature range, comparable with those reported in literatures [4,20–29].

Next, we annealed the as-grown samples in oxygen *ex situ*. The result is shown in Figure 1b. It can be seen that annealing significantly reduces the conductivity of all the heterointerfaces. Only in $LaAlO_3/SrTiO_3$ (squares) and LSAT/$SrTiO_3$ (diamonds) the metallic behavior is fairly robust to an annealing at 600 °C. The metallic behavior of $LaAlO_3(111)/SrTiO_3$, $LaAlO_3(110)/SrTiO_3$, $LaTiO_3/SrTiO_3$, and $\gamma$-$Al_2O_3/SrTiO_3$ can fairly survive an annealing at 300~400 °C. The conductivity of all others, after annealing at 300 °C or lower, is too poor to be measured in our cryostat system (thus not shown). To make a close comparison, in Figure 1c we grouped the heterointerfaces into four classes (polar band insulators, polar Mott insulators, non-polar band insulators, and amorphous oxides), and summarized their sheet carrier density, $n_{sheet}$, in all cases. Here $n_{sheet}$ is estimated from $R_{sheet}$ at 300 K [33], using a simple empirical relation, $n_{sheet} = 1/e(\text{mobility}\times R_{sheet})$, where $e$ is the unit charge, and assuming a constant mobility of 6 $cm^2V^{-1}s^{-1}$, a typical value for $SrTiO_3$ at room temperature [13,35].

As expected, the as-grown samples without annealing have much higher $n_{sheet}$ than that of the annealed ones, which can be attributed to the existence of large amount of oxygen vacancies. The fact that so many heterointerfaces with polar capping layers ($LaGaO_3$, $GdTiO_3$, and *etc.*) become insulating after annealing suggests that polar discontinuity and electronic reconstruction is not a general mechanism because it is less likely that the moderate annealing condition used here (300 °C or lower) can fully destroy the polar arrangement in the capping materials. Because the main effect of annealing is to fill oxygen vacancies, in both the $SrTiO_3$ substrate and the capping oxide film, we conclude that oxygen vacancies must play a key role in the formation of interfacial conductivity. In addition, because all the heterointerfaces have the same $SrTiO_3$ substrates (except for the different crystal orientations in $LaAlO_3(111)/SrTiO_3$ and $LaAlO_3(110)/SrTiO_3$), the extremely large scattering of interfacial conductivity after annealing implies that the oxygen vacancies in the capping oxide films, whose stability can vary largely for different capping materials, should play an important role on the formation of interfacial conductivity. Yu *et al.* theoretically proposed that surface oxygen vacancies can provide electrons to the interface of $LaAlO_3/SrTiO_3$ [7]. However, for any real samples, their surfaces are almost certainly covered by water-related adsorption due to the ubiquitous environmental water. Furthermore, it is well known that water chemistry can occur on oxide surface either at defects [36] or metal ions [37], but only oxygen vacancies will be significantly affected by annealing. Therefore, water chemistry at surface oxygen vacancies, rather than the oxygen vacancies themselves, should be more likely to account for the interfacial electrons in real situations.

**B. Effect of surface treatment with water.**

To examine the possibility of water chemistry at surface oxygen vacancies, we treated various heterointerfaces with deionized (DI) water. The results are summarized in Figure 2 (Similar results were also obtained by treating the heterointerfaces with acetone and ethanol, Figure S16 [33]). Note that the surface treatment used here is only a post tuning of surface chemistry because all of our samples have already been exposed to ambient environment before treating, and their surfaces should have already been covered with plenty of adsorbates, especially water related species [38].

Remarkably, a close relation between the effect of surface treatment and the amount of surface oxygen vacancies is observed. For the as-grown (without annealing) samples, the $R_{sheet}$ only change slightly after surface treatments (Figure 2a, black squares.) Much larger changes in $R_{sheet}$ were observed for the annealed samples (Figure 2a, blue circles and red triangles), if the annealed samples weren't fully insulating. The tuning effect can be better evaluated from $\Delta n_{sheet}$ (Figure 2b) and $\Delta R_{sheet}$ ratio (Figure 2c). Taking $LaAlO_3/SrTiO_3$ as an example: the $\Delta R_{sheet}$ ratio is tiny for the as-grown and 300 °C-annealed samples, but much larger for the 400 & 600 °C-annealed samples (Figure 2c). We emphasize that even the absolute change, $\Delta n_{sheet}$, is also much smaller for the as-grown $LaAlO_3/SrTiO_3$ (Figure 2b), compared with those annealed above 400 °C. A careful analysis of all the data shown in Figure 2 show that it is generally true that the conductivity of a heterointerface (except for these with amorphous oxides) is much more sensitive to the surface treatment when it is in a status of less surface oxygen vacancies. This statement is true even for the samples without annealing. For example, we found that the resistance of the as-grown $CaZrO_3/SrTiO_3$ is instable and increases fairly quickly in ambient environment (not shown), indicating that oxygen can be filled into $CaZrO_3$ at this condition; accordingly, the as-grown $CaZrO_3/SrTiO_3$ has the largest $\Delta R_{sheet}$ ratio among all the heterointerfaces (Figure 2c, open square). As will be discussed below, this behavior can be understood by that when the amount of surface oxygen vacancies is low, less surface water chemistry can occur.

As shown in Figure 2, it is interesting to see that almost all the heterointerfaces with a *crystalline* capping layer, no matter the layer is polar band, polar Mott, or non-polar band insulator, can show a significantly enhancement in the conductivity after surface treatments, if the amount of their surface oxygen vacancies are not too high. By contrast, when the capping layer is *amorphous*, the conductivity decreases slightly after surface treatments. This result is not surprising because much more oxygen vacancies are expected on the surface of the amorphous oxides. In addition, we notice that in many cases, even the annealed insulting samples can be recovered to a conducting status by a surface treatment. If a naive gauge of robustness to oxygen annealing is defined as, less increase in $R_{sheet}$ > insulating but recoverable by surface treatment > insulating and irrecoverable, from the data shown in Figures 1b & 2a, a rough rank for all these heterointerfaces can be constructed as shown in Figure 3a.

**C. AP-XPS measurements.**

To obtain further insights of surface water chemistry, AP-XPS measurements were performed. Figure 4 shows the results of a $LaAlO_3/SrTiO_3$ (c-LAO) sample and an amorphous-$LaAlO_3/SrTiO_3$ (a-LAO) sample with namely the same capping-layer-thickness of 4 nm. Note that c-LAO corresponds to the $LaAlO_3/SrTiO_3$ sample annealed *in situ*, as the one labeled by the star symbol in Figure 2c ($^*LaAlO_3/SrTiO_3$). The two samples were transferred simultaneously into the AP-XPS chamber through air, and measured as in. As shown in Figure 4a, the O 1s spectra of c-LAO (circles) shows a strong hydroxyl (OH) component in addition to the oxide component, consistent with previous studies [11,14,16,39], confirming that water chemistry occurs on the surface of $LaAlO_3/SrTiO_3$. The OH component in the spectra of a-LAO (triangles) is much stronger, indicating a much stronger water chemistry, consistent with the expectation that there are more oxygen vacancies in a-LAO. After applying a water vapor of 0.3 mbar during the AP-XPS measurements, the OH component in both samples increases a bit (Figures 4b & 4c), indicating an increase in the surface water chemistry. An O 1s shift of ~0.2 eV to higher binding energy (BE) was observed in c-LAO (Figure 4b). Similar BE shifts were also observed in the core levels of La 4d and Al 2p, but not in Sr 3d and Ti 2p (Figure 4d, black squares). In another word, only the core levels in the $LaAlO_3$ side (La 4d, Al 2p, and O 1s) have a concomitant BE shift. This kind of distinct BE shift is very similar to that observed for the surface-hydrogen-adsorption induced electron transfer [16,33], strongly supporting that surface water chemistry induces electron transfer in c-LAO. By contrast, applying water vapor does not cause any detectable BE shift in a-LAO (Figures 4c & 4d (blue circles)), indicating a lack of electron transfer. These observations consist well with the surface treatment result shown in Figure 2 which shows that a surface treatment of water can significantly increase the conductivity of c-LAO ($^{*}LaAlO_3/SrTiO_3$), but not for a-LAO.

Another striking observation is that the BE of O 1s (oxide) of a-LAO, compared with that of c-LAO, shifts to higher BE of ~1.2 eV (Figure 4a). Again, the BE shifts of similar values were observed in the core levels of La 4d and Al 2p, but not in Sr 3d and Ti 2p (Figure 4d, red circles). This observation indicates that much more electrons had been transferred in the a-LAO than the c-LAO. It is reasonable because significant surface water chemistry should have already occurred on both the a-LAO and the c-LAO, due to the ubiquitous water (even without intentionally applying surface treatment or water vapor), and stronger water chemistry is expected in a-LAO because it has more surface oxygen vacancies.

## IV. DISCUSSION

In the following we show how all the present observations can be consistently understood in a simple electrostatic consideration based on electron transfer from water chemistry at surface oxygen vacancies. For simplicity, we represent the effect of surface water chemistry as providing surface shallow levels (SSL) that can supply electrons to the interface. In the present discussion we temporarily ignore the influence of bulk oxygen vacancies in $SrTiO_3$ substrate, the possible SSL in the bulk of the capping oxide layers, and other complexities. We emphasize that these factors only affect the magnitude of interfacial conductivity, and can coexist with the present scenario (a discussion including these factors can be found in the Table III, and the (SM) [33]).

Figure 5a shows an ideal band alignment of an imaginary situation when a capping oxide layer is contacting with a $SrTiO_3$ substrate, but in a moment no electron transfer has occurred. The whole bands (including conduction band, valence band, and core levels) in the capping oxide are either tilting up or flat, depending on the polar nature of the capping layer. $E_{\mathrm{drive}}$ is defined as the energy difference between SSL and the Fermi level ($E_F$) in $SrTiO_3$. Obviously, if $E_{\mathrm{drive}} > 0$, the electrons initially existing in SSL can transfer into the interface, forming interfacial conductivity. This electron transfer will create an electrical field that will reduce the potential in the capping oxide layer (Figure 5b). Either exhausting of all the available SSL or $E_{\mathrm{drive}}=0$ (after transferring electrons) will stop the electron transfer process.

The flat band behavior of $LaAlO_3/SrTiO_3$ [39,40] observed previously can be understood by that electron transfer from ambient water chemistry (always exists for all the real samples) happens to cancel out the built-in potential in $LaAlO_3$. It also indicates that in equilibrium SSL are exhausting but $E_{\mathrm{drive}}>0$. Therefore, surface treatments or applying water vapor can increase SSL by increasing water chemistry, resulting in a further electron transfer, which leads to an increase in conductivity (Figure 2) and BE shift (Figures 4b & 4d). Similar explanation can be extended to other heterointerfaces that consist of crystalline capping layers. However, the situation is different for the amorphous capping oxides. Because there is no polarity in amorphous oxides, without electron transfer the initial bands in amorphous oxides should be flat. As shown in Figure 4a, in amorphous oxides there are huge amount of water chemistry that can provide abundant SSL. Therefore, electrons can transfer from SSL to the interface until $E_{drive}$=0. The transferred electrons tilt downwards the whole bands in the amorphous $LaAlO_3$ layer, causing a large BE shift (Figures 4a & 4d). Because $E_{drive} = 0$, further water chemistry cannot cause electron transfer any more, explaining the lacking of BE shift after applying water vapor (Figure 4c). Therefore, for heterointerfaces with amorphous oxides, the conductivity is expected to be insensitive to surface treatments. In fact, we observed a slight decrease in conductivity (Figure 2). Tentatively, we attribute it to a de-protonation effect [14] or a tiny change in work function due to the alteration of surface adsorption.

Following the same argument, the surface of the as-grown samples, compared with the annealed samples, have much more oxygen vacancies that can provide more SSL by ambient water chemistry. Consequently, it will result in a larger surface-to-interface electron transfer and a smaller $E_{drive}$, which explains why the as-grown heterointerfaces are less sensitive to the surface treatment (Figure 2). On the other side, annealing in oxygen will reduce the amount of surface oxygen vacancies, and equivalently reduce SSL, which will reduce the amount of transferred electrons and subsequently increase $E_{drive}$, which explains why a heterointerface is much more sensitive to the surface treatment when it is in a status of less surface oxygen vacancies (Figure 2). Too severe annealing will remove all the available SSL, and thus make samples insulating. A previous theoretical calculation showed that the surface oxygen vacancies in $LaAlO_3/SrTiO_3$ are energetically stable [7], which can explain why the metallic behavior of $LaAlO_3/SrTiO_3$ (LSAT/$SrTiO_3$) can survive a severe annealing. We caution that if water chemistry occurs on surface Al ions [39] instead, a strong robustness to annealing is also expected because annealing will not change the amount of surface Al ions. However, even in this case water chemistry at the surface oxygen vacancies should also have a significant contribution because both the conductivity and the sensitivity of surface treatment of $LaAlO_3/SrTiO_3$ and LSAT/$SrTiO_3$ show a strong dependence on annealing conditions.

In addition, we note that many important previous observations in $LaAlO_3/SrTiO_3$ system can be well explained within the present scenario. The existence of a critical thickness of $LaAlO_3$ [35] and the dependence on the termination of $SrTiO_3$ [5] can be explained by that the formation of *surface* oxygen vacancies [7] depends on the thickness of $LaAlO_3$ and the termination of $SrTiO_3$. That off-stoichiometry (Al-rich) of $LaAlO_3$ enhances conductivity [18,19] can be explained by that water chemistry is more favorable for particular kinds of surface defects. More observations and their explanations with the present scenario are summarized in the Table 2.

Finally, we crudely estimated the value of $E_{drive}$ by the energy difference between the conduction band minimum at the surface of capping oxides and that of $SrTiO_3$ (SM [33]). To take into account of the contribution from polar layer, we arbitrarily added 1 eV, based on a previous measurement on $LaCrO_3/SrTiO_3$ [41], to all the polar oxides. The trend of such estimated $E_{drive}$ agrees fairly well with the trend of robustness to oxygen annealing (Figure 3b). The only large scattering is in γ-$Al_2O_3$, but it is not surprising since γ-$Al_2O_3$ has a completely different lattice structure from all other perovskite oxides. The coincidence of $E_{drive}$ and the robustness to oxygen annealing implies that the stability of surface oxygen vacancies are determined by the intrinsic properties of the capping oxides.

## V. CONCLUSION

Our present study suggests that water chemistry at surface oxygen vacancies is a common mechanism of the interfacial conductivity of all kinds of complex oxide heterointerfaces; the role of the intrinsic properties of the capping oxides is mainly on determining the stability of surface oxygen vacancies. This scenario shares the same electronic reconstruction feature with the polar discontinuity related mechanisms [5,7]. It differs from the latters in two key points: (1) The origin of interfacial electrons is surface chemistry rather than valence band of surface capping oxides, or surface oxygen vacancies themselves. (2) A potential difference, $E_{drive}$, rather than the built-in potential from polar discontinuity, is the driving force for electron transfer. In this scenario, polar discontinuity is not indispensable for the interfacial conductivity, although it improves $E_{drive}$ and facilitates electron transfer. More comparisons between the present and previous mechanisms are summarized in the Table 3.

## ACKNOWLEDGEMENTS

The authors would like to acknowledge Scott Chambers at PNNL for very helpful discussion. This work was supported by the National Key R&D Program of the MOST of China (Grants No. 2016YFA0300204 and No. 2017YFA0303002), Thousand Youth Talents Plan, National Natural Science Foundation of China (11227902), and the Fundamental Research Funds for the Central Universities.

**Figure captions.**

FIG.1. Transport characterization. Temperature dependence of $R_{sheet}$ of (a) the as-grown heterointerfaces and (b) after *ex-situ* annealing. Capping materials and annealing temperatures are as labeled. “a-” stands for amorphous. The symbol “*” means that the corresponding sample was annealed at 600 °C *in situ*. (c) The estimated $n_{sheet}$ at 300 K of various as-grown and annealed heterointerfaces. $LaAlO_3$(110)/$SrTiO_3$ is regarded as “non-polar” because there is no polar discontinuity at its interface. The lines are the guides for the eye.

FIG.2. Effect of surface treatments by DI water. (a) $R_{sheet}$ of the as-grown and annealed heterointerfaces before and after surface treatments. (b) $\Delta n_{sheet}$, defined as [$n_{sheet}$ (after)-$n_{sheet}$ (before)]. (c) $\Delta R_{sheet}$ ratio, defined as [$R_{sheet}$ (after)-$R_{sheet}$ (before)]/$R_{sheet}$ (before). For comparison, the result of the *in situ* annealed samples is also shown in (c). All data were measured at room temperature. The lines are the guides for the eye.

FIG. 3. (a) A rough rank for the robustness of the conductivity of various heterointerfaces to annealing in oxygen, constructed from the data shown in Figures 1(b) & 2(a). (b) The estimated $E_{drive}$ *vs* robustness to oxygen annealing.

FIG. 4. AP-XPS measurement of crystalline and amorphous $LaAlO_3$/$SrTiO_3$ heterostructures. (a) O1s spectra of the “as in” c-LAO and a-LAO. The symbols represent the raw data. The thick lines are the envelopes of the spectra. The grey dotted lines are the fitted components of the spectra. The spectra have been rescaled to have the same intensity of the O 1s(oxide) component, as well as in (b &c). (b) O1s spectra of the c-LAO before and after exposure to water vapor. (c) O1s spectra of the a-LAO before and after exposure to water vapor. (d) Comparison of core level positions in three different conditions as annotated in the panel.

FIG. 5. Schematic drawing of electron transfer and band alignment. (a) Before and (b) after transferring electrons. $E_{drive}$ is not strictly defined since SSL will distribute in an energy region rather than a single energy level. BE measured by AP-XPS is the energy difference between core level and $E_F$. A downward tilting of bands will result in an increase in BE. CBM and VBM stand for conduction band minimum and valence band maximum, respectively.

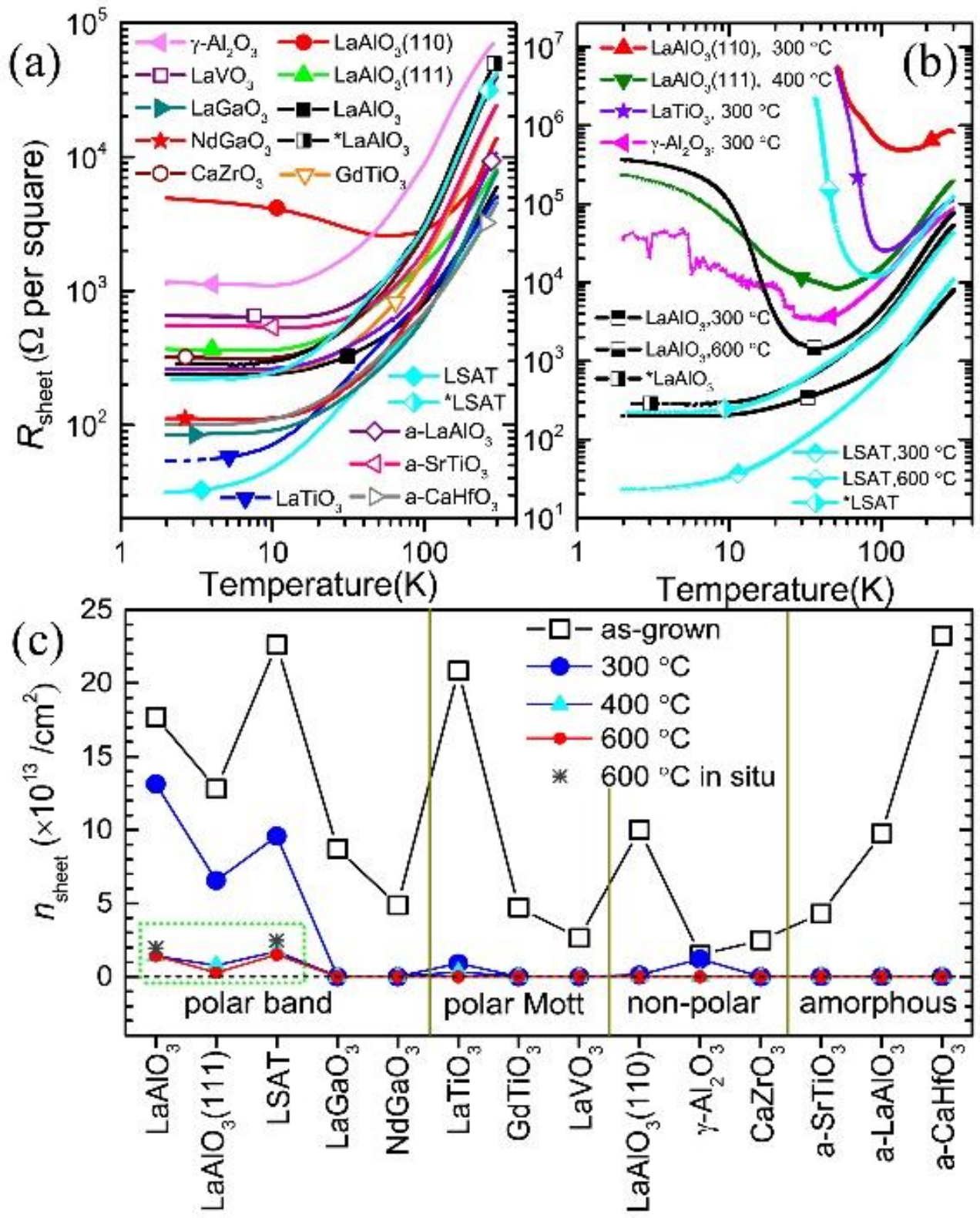

**FIG. 1 by Zhang *et al.***

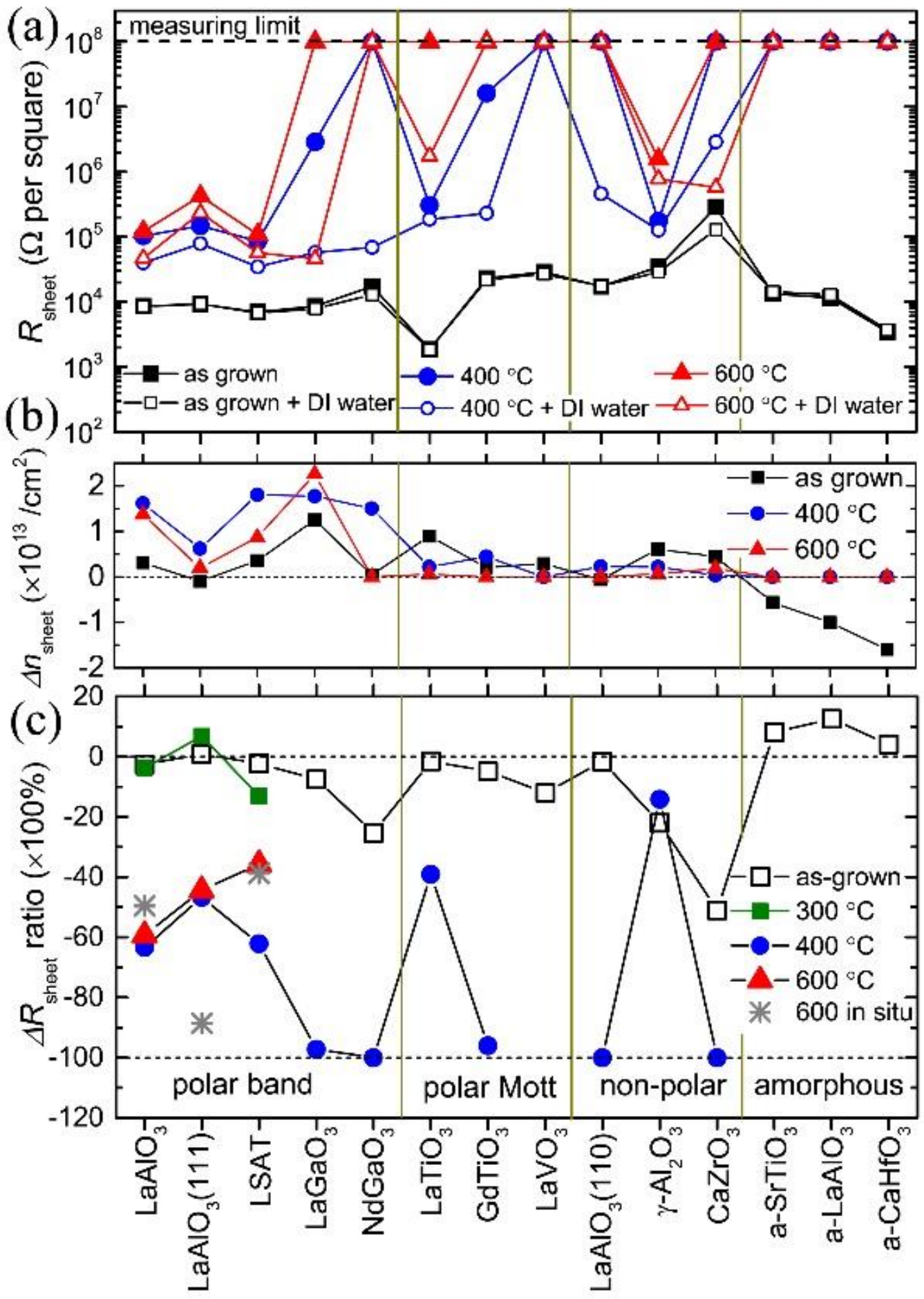

**FIG. 2 by Zhang *et al.***

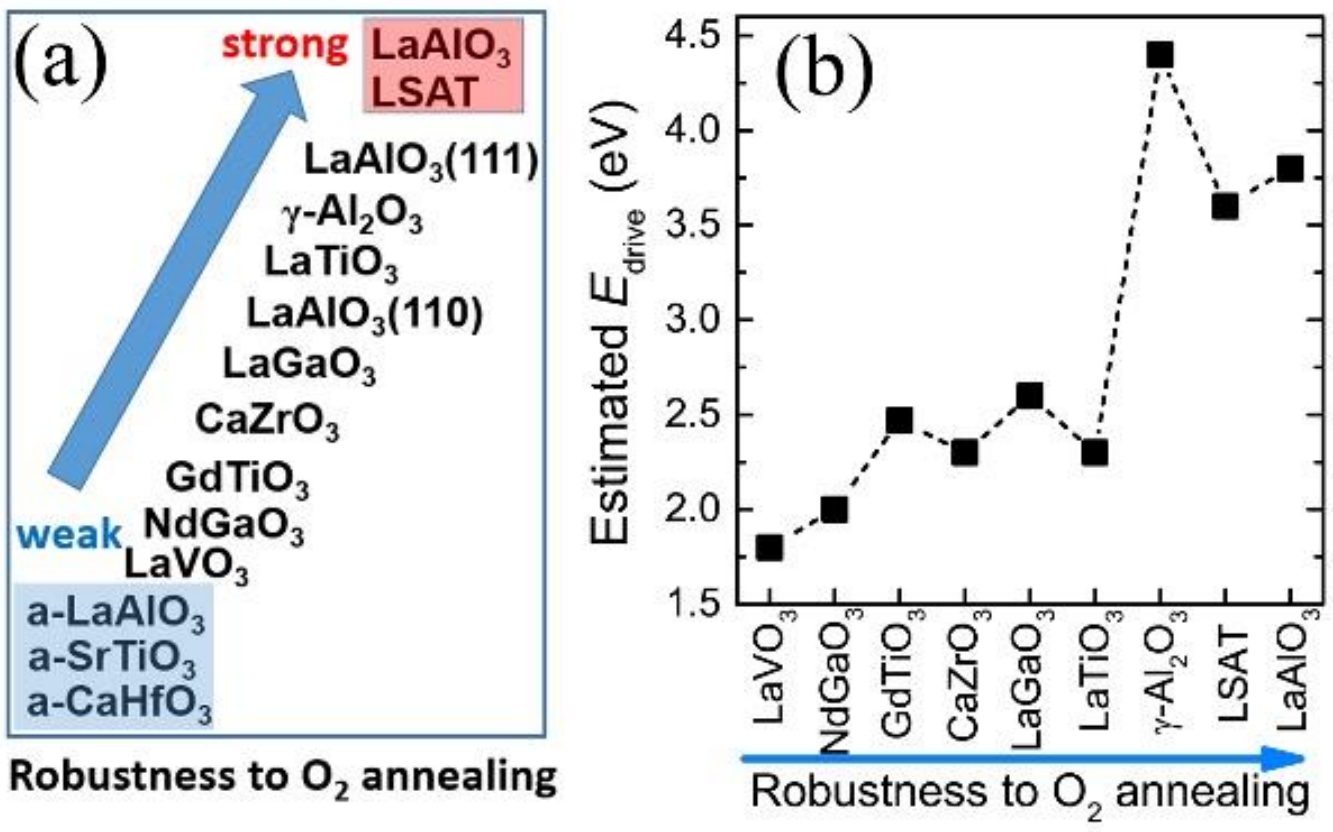

**FIG. 3 by Zhang *et al.***

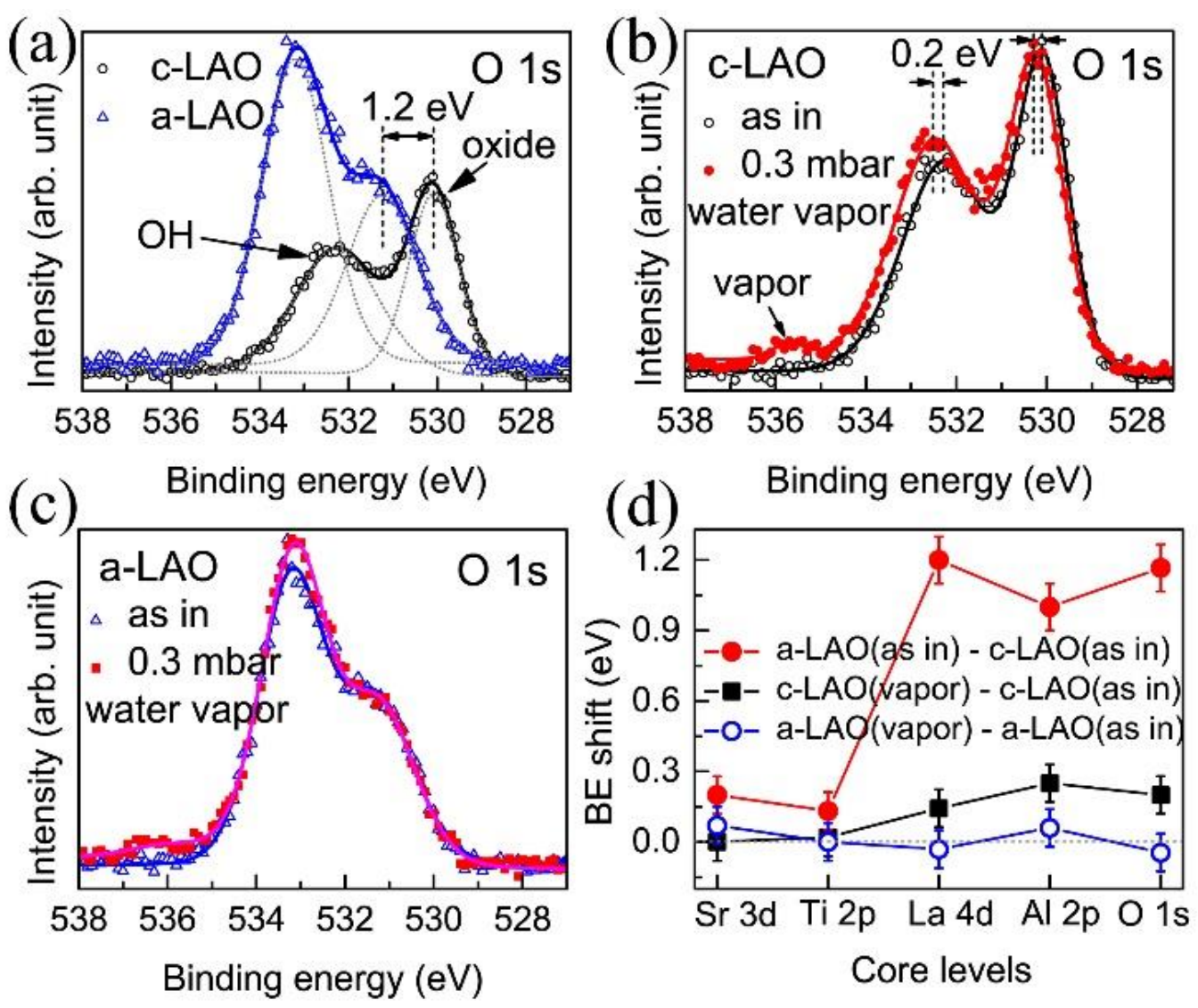

**FIG. 4. by Zhang *et al.***

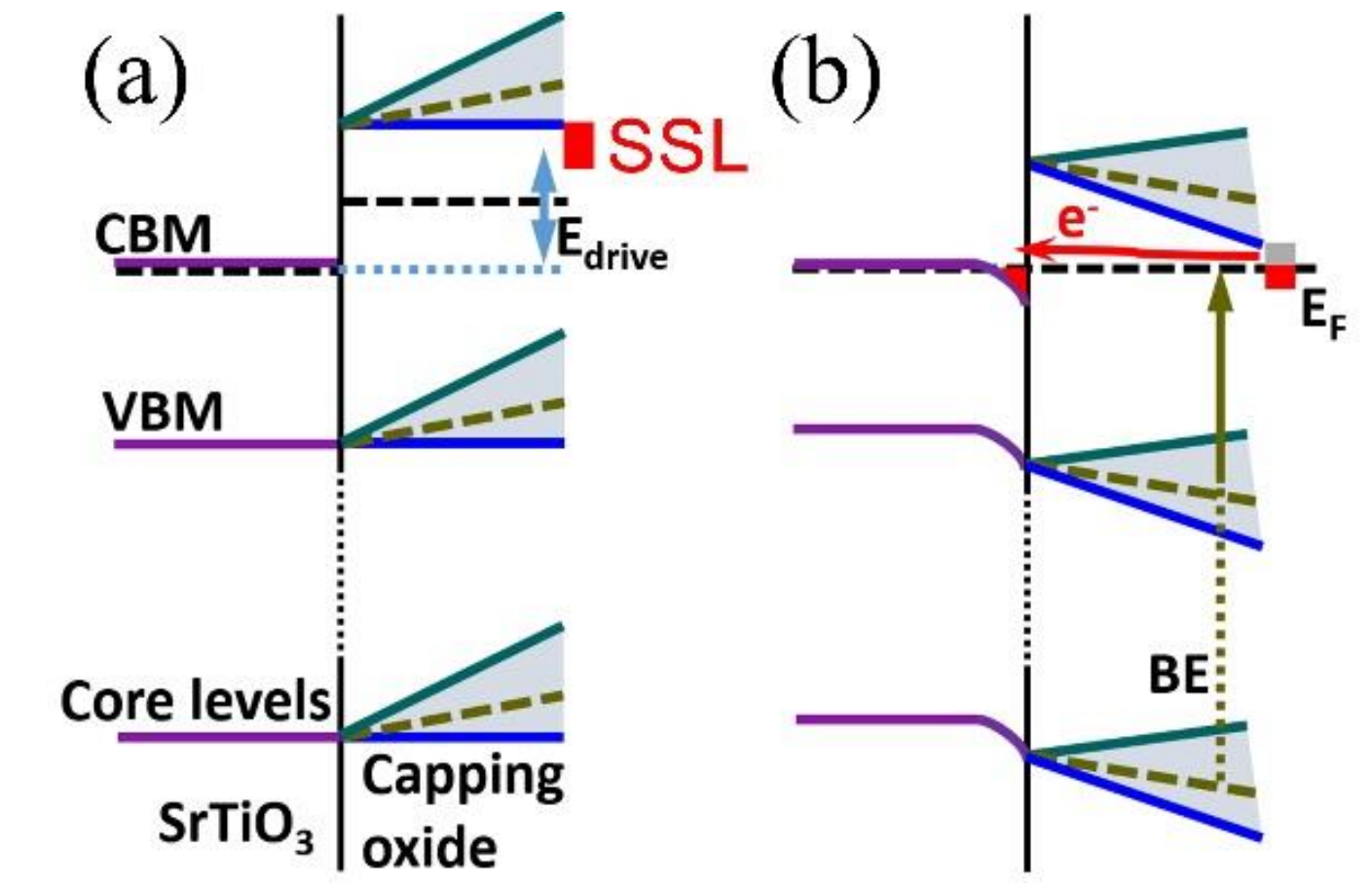

**FIG. 5 by Zhang *et al.***

Table 1. Details of growth parameters for different capping materials grown on $SrTiO_3$.

| **Samples** | **T (growth) (°C)** | **$P(O_2)$ (mbar)** | **Laser fluence ($Jcm^{-2}$)** | **In situ post-annealing** | **thickness** |
|---|---|---|---|---|---|
| $LaAlO_3$ | 800 | $10^{-5}$ | 0.8 | No | 10 uc |
| *$LaAlO_3$ | 800 | $10^{-5}$ | 0.8 | 600 °C, 200 mbar $O_2$, 1 hour | 10 uc |
| $LaAlO_3$(111) | 850 | $10^{-4}$ | 1.4 | No | 500 pulses (15~18 uc) |
| *$LaAlO_3$(111) | 850 | $10^{-4}$ | 1.4 | 600 °C, 200 mbar $O_2$, 1 hour | 500 pulses (15~18 uc) |
| LSAT | 800 | $10^{-3}$ | 1.8 | No | 10 uc |
| *LSAT | 800 | $10^{-3}$ | 1.8 | 600 °C, 200 mbar $O_2$, 1 hour | 10 uc |
| $LaGaO_3$ | 800 | $10^{-4}$ | 2.0 | No | 250 pulses (9~11 uc) |
| $NdGaO_3$ | 750 | $10^{-3}$ | 2.0 | No | 250 pulses (9~11 uc) |
| $LaTiO_3$ | 625 | $7\times10^{-6}$ | 0.9 | No | 300 pulses (8~10 uc) |
| $GdTiO_3$ | 650 | $10^{-4}$ | 2.0 | No | 500 pulses |
| $LaVO_3$ | 600 | $7\times10^{-6}$ | 2.5 | No | 200 pulses (9~11 uc) |
| $LaAlO_3$(110) | 720 | $10^{-4}$ | 1.4 | No | 500 pulses (15~18 uc) |
| $\gamma$-$Al_2O_3$ | 600 | $10^{-4}$ | 1.2 | No | 500 pulses (2~3 uc) |
| $CaZrO_3$ | 600 | $10^{-4}$ | 1.5 | No | 250 pulses (10~12 uc) |
| a-$SrTiO_3$ | room temp | $5\times10^{-6}$ | 1.0 | No | 500 pulses (~4 nm) |
| a-$LaAlO_3$ | room temp | $5\times10^{-6}$ | 1.0 | No | 500 pulses (~3 nm) |
| a-$CaHfO_3$ | room temp | $10^{-4}$ | 3.0 | No | 140 pulses (~4 nm) |

Table 2. Possible explanation to a few important phenomena.

| Phenomenon | Explanation within the present model |
|---|---|
| (1)$TiO_2$-terminated $LaAlO_3/SrTiO_3$ is conducting; SrO-terminated $LaAlO_3/SrTiO_3$ is insulating [4]. | *Surface* oxygen vacancies are thermodynamically *favorable* in $TiO_2$-terminated $LaAlO_3/SrTiO_3$, but *unfavorable* in SrO-terminated $LaAlO_3/SrTiO_3$. (Nat. Commun. **5**, 5118, (2014) by Liping Yu and Alex Zunger) [7] |
| (2) There is a critical thickness in $LaAlO_3/SrTiO_3$ [35]. | A critical thickness is needed for the presence of *surface* oxygen vacancies. (Nat. Commun. **5**, 5118, (2014) by Liping Yu and Alex Zunger). [7] |
| (3) Properties of $LaAlO_3/SrTiO_3$ samples vary largely from lab to lab. | In most studies, *surface* oxygen vacancies, which can be affected by many growth conditions, were not well controlled, and subsequently affect surface adsorption/chemistry. |
| (4) XPS measurements showed that in $LaAlO_3/SrTiO_3$ the bands of $LaAlO_3$ are nearly flat: no detectable remaining built-in potential [39,40]. | Surface chemistry provides electrons that have much shallower energy levels than the valence band of $LaAlO_3$. These electrons can readily transfer into the interface, producing an additional electrical field that cancels out the built-in potential. |
| (5) So many heterointerfaces are conducting even when the capping layer is non-polar. | Electrons can transfer from surface to interface once surface chemistry generates electrons with energy levels higher than the $E_F$ (also the conduction band bottom) of STO.<br><br>A *polar* capping layer is helpful because it can stabilize the surface oxygen vacancies, facilitate surface chemistry, and increase the energy levels of surface electrons. However, it is *not a prerequisite* for electron transfer. |
| (6) So many conducting heterointerfaces are not robust against annealing in oxygen. | Annealing in oxygen can remove *surface* oxygen vacancies, which equivalently reduce the surface adsorption/chemistry. |
| (7) In contrast to $LaAlO_3/SrTiO_3$, $LaFeO_3/SrTiO_3$ and $LaCrO_3/SrTiO_3$ are typically insulating while a remaining built-in potential nearly 1 eV was indeed observed within $LaFeO_3$ and $LaCrO_3$ [PRL**107**, 206802 (2011) by Scott Chambers *et* | Compared with $LaAlO_3$ and many other capping materials, the conduction bands of $LaFeO_3$ and $LaCrO_3$ are much deeper and very close to that of $SrTiO_3$ (PRL**107**, 206802 (2011) and PRL **117**, 226802 (2016), by Scott Chambers *et al.*); in addition, Fe and Cr are multi-valence elements. Therefore, electrons prefer to transfer into $LaFeO_3$ and $LaCrO_3$ *first* when the conditions for electron transfer are met (by increasing film thickness since both $LaFeO_3$ and |
| *al.*] [41]. | $LaCrO_3$ are polar). The fact that there is still a remaining built-in potential indicates that this amount of potential is needed to raise the energy levels of surface electrons to fulfill the conditions for transferring (into $LaFeO_3$ and $LaCrO_3$). |
| (8) Off-stoichiometry is crucial for interfacial conductivity [Nat. Commun.**4**, 2351 (2013) by D. G. Schlom *et al.* [18] & PRL**110**, 196804 (2013) by L. W. Martin *et al.* [19]] | Surface defects, which are crucial for surface adsorption/chemistry, can be significantly affected by off-stoichiometry. |
| (9) In XPS measurement, with increasing the thickness of $LaAlO_3$, both $TiO_2$-terminated $LaAlO_3/SrTiO_3$ and SrO-terminated $LaAlO_3/SrTiO_3$ show a similar increase in binding energy of Al 2p core level [PRB **84**, 245124(2011) by M. Takizawa *et al.*] [42]. | This thickness-dependent BE shift should not be attributed to any remaining built-in potential in LAO. It comes from surface-to-interface electron transfer. Before electron transfer happens, the $LaAlO_3$ bands of both $TiO_2$-terminated $LaAlO_3/SrTiO_3$ and SrO-terminated $LaAlO_3/SrTiO_3$ are actually nearly *flat*. The strong X-ray beam generates additional surface oxygen vacancies [Nature **469**, 189 (2011) by A.F. Santander-Syro *et al.* [43]; Nat. Mat. 10, 114 (2011) by Z. X. Shen *et al.* [44]] which can adsorb the remaining water or hydrogen in the measurement chamber [PRB **92**, 195422 (2015) by P. Scheiderer *et al.*], resulting in electron transfer. Given a constant amount of transferred electrons, the induced decrease in potential (or increase in BE) is linearly proportional to the thickness of $LaAlO_3$. This explains the $LaAlO_3$-thickness-dependent BE shift in both kinds of heterointerfaces. |

Table 3. A brief comparison with other mechanisms.

| Other mechanisms | The present model |
|---|---|
| Polar discontinuity and electronic reconstruction | Difference: (a) in the present model the driving force for electronic reconstruction is the energy difference in chemical potential, rather than the built-in potential due to polar discontinuity; (b) the transferred electrons are from surface chemistry, rather than the valence band of $LaAlO_3$.<br>Relation: Most arguments previously made based on the "polar discontinuity and electronic reconstruction" are still basically correct in the present model, with a benefit that the present model is more general and practical. |
| Polarity-induced defect mechanism | As $LaAlO_3/SrTiO_3$ is concerned, in the present model the transferred electrons are from water chemistry controlled by surface oxygen vacancies, rather than the surface oxygen vacancies themselves.<br>The present model is more general and practical, because water is ubiquitous in ambient environment and tends to react with surface oxygen vacancies. |
| Surface redox reaction or hydrogen adsorption | Similar idea, but the present model is more general. |
| Oxygen vacancies in bulk $SrTiO_3$ | (a) They can coexist with the present model.<br>(b) Most previous studies attribute the extra conduction of $LaAlO_3/SrTiO_3$ samples prepared in low oxygen pressure to the extrinsic mechanism of oxygen vacancies in bulk $SrTiO_3$ completely; the present model suggests that in this case a significant part of the extra conduction should come from surface adsorption due to the large amount of surface oxygen vacancies. |
| Interface mixture, defects, and *etc*. | If any, they can coexist with the present model. |

## Supplemental Material for

# Origin of interfacial conductivity at complex oxide heterointerfaces: possibility of electron transfer from water chemistry at surface oxygen vacancies

Meng Zhang[1,2], Zheng Chen[1,2], Baohua Mao[3], Qingtian Li[3], Hai Bo[1,2], Tianshuang Ren[1,2], Pimo He[1,2], Zhi Liu,[3,4] and Yanwu Xie[1,2*]

[1]Department of Physics, Zhejiang University, Hangzhou 310027, China

[2]Collaborative Innovation Center of Advanced Microstructures, Nanjing University, Nanjing 210093, China

[3]State Key Laboratory of Functional Materials for Informatics, Shanghai Institute of Microsystem and Information Technology, Chinese Academy of Sciences, Shanghai 200050, China

[4]School of Physical Science and Technology, ShanghaiTech University, Shanghai 200031, China

[*]Correspondence and requests for materials should be addressed to Y. X. (email: ywxie@zju.edu.cn)

## 1. Structural characterization.

RHEED and atomic force microscopy characterization of typical samples were shown in the following figures.

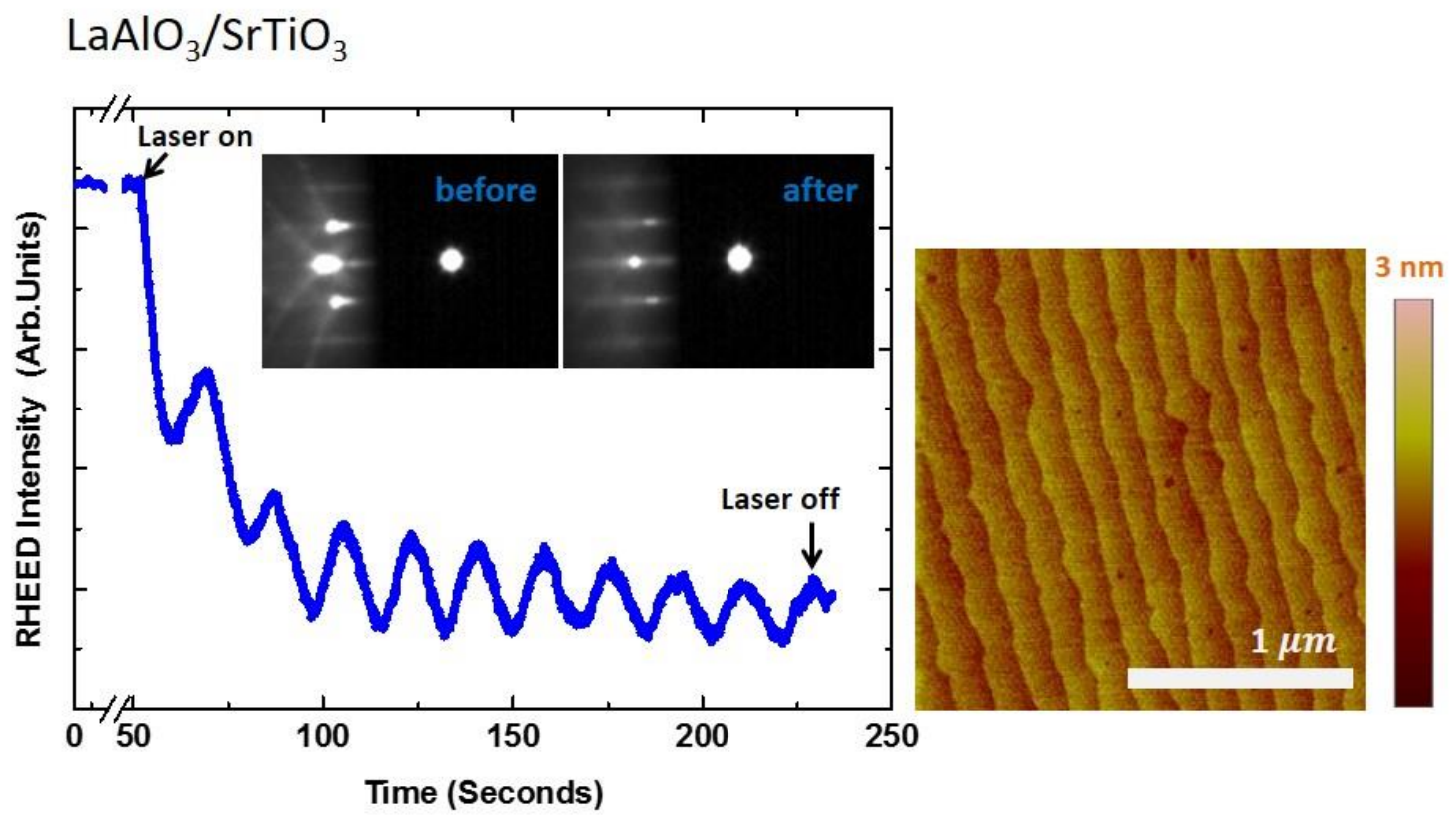

Figure S1. $LaAlO_3/SrTiO_3$

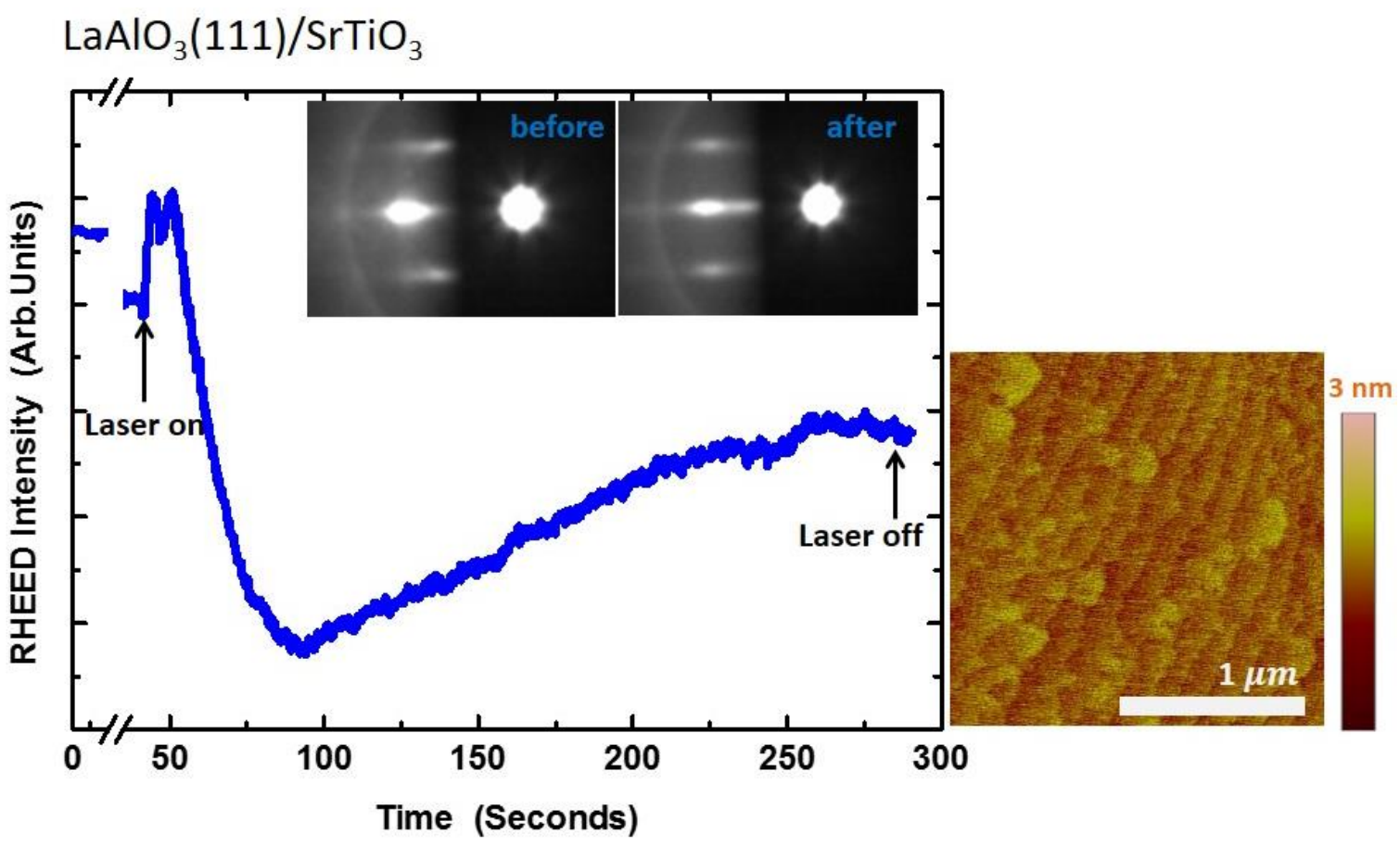

Figure S2. $LaAlO_3$(111)/$SrTiO_3$

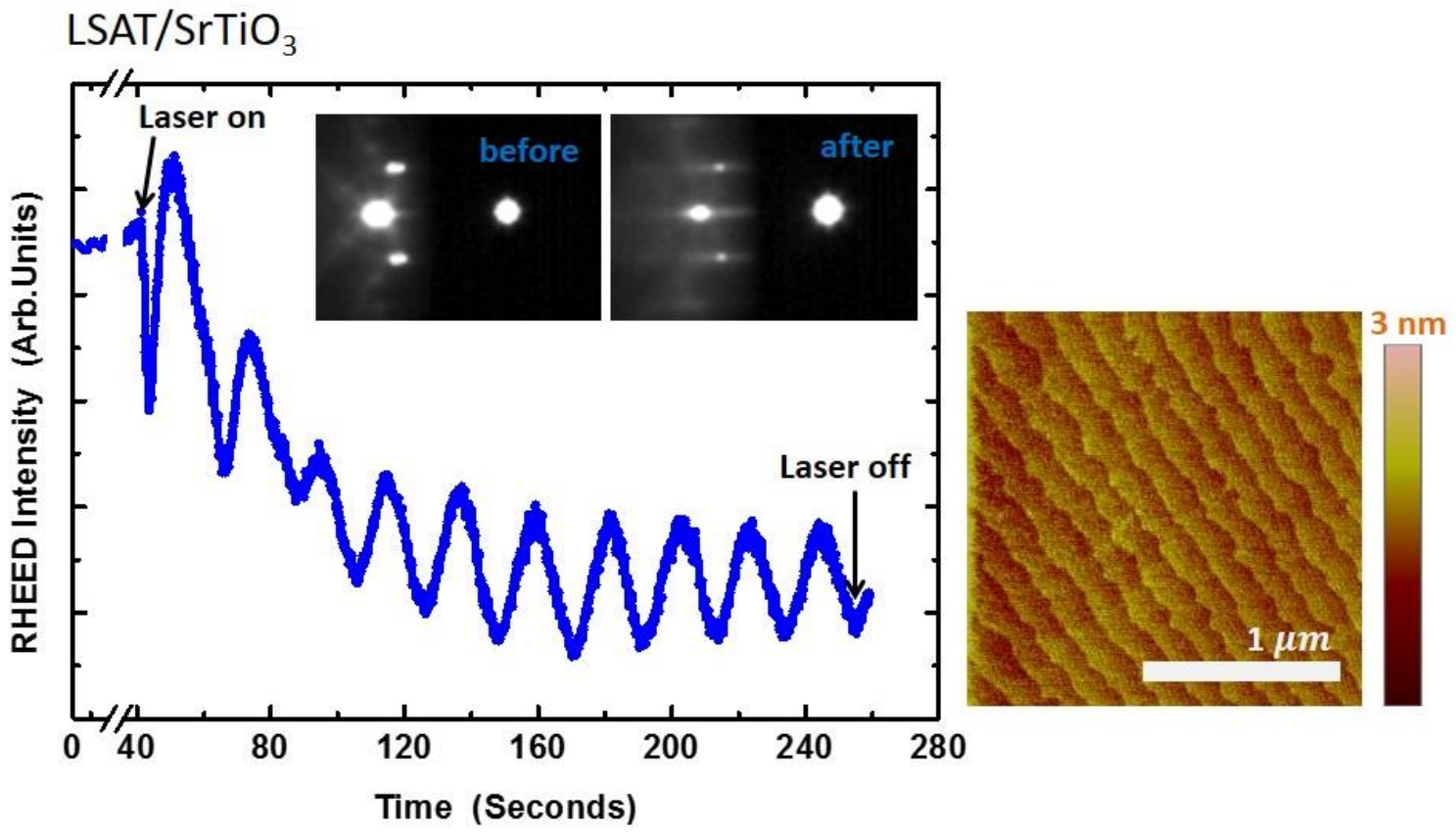

Figure S3. LSAT/$SrTiO_3$

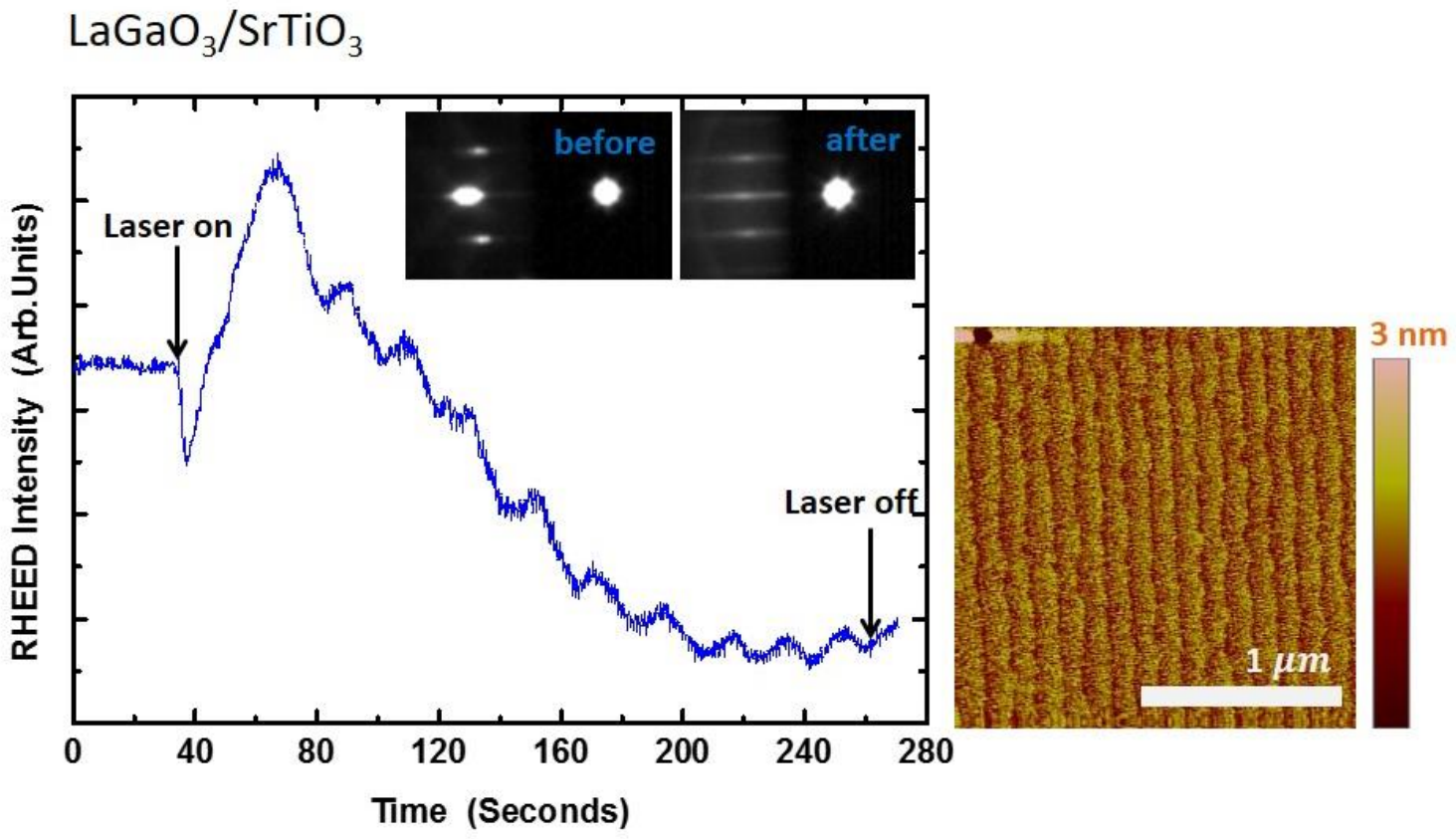

Figure S4. $LaGaO_3$/$SrTiO_3$

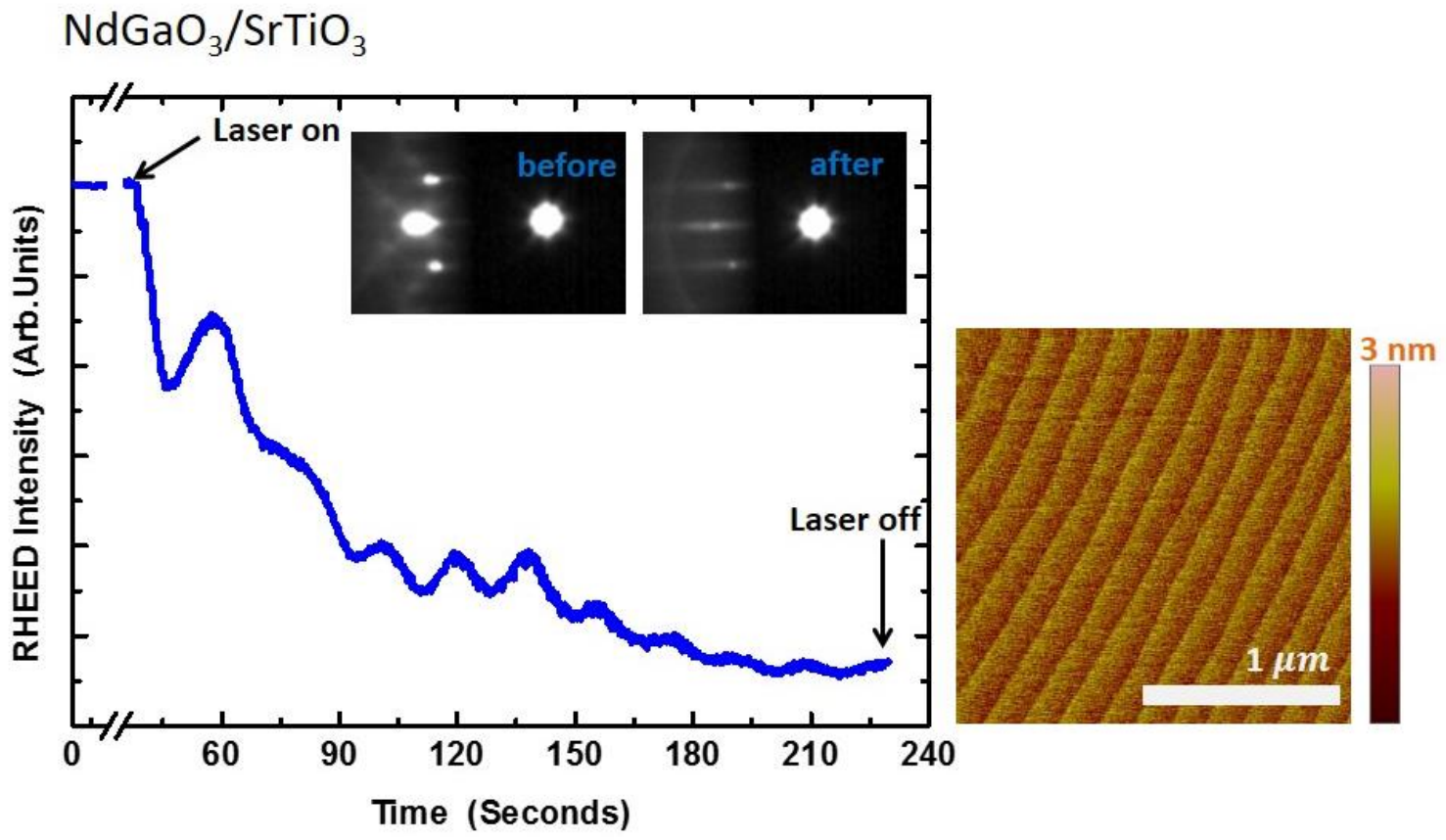

Figure S5. $NdGaO_3/SrTiO_3$

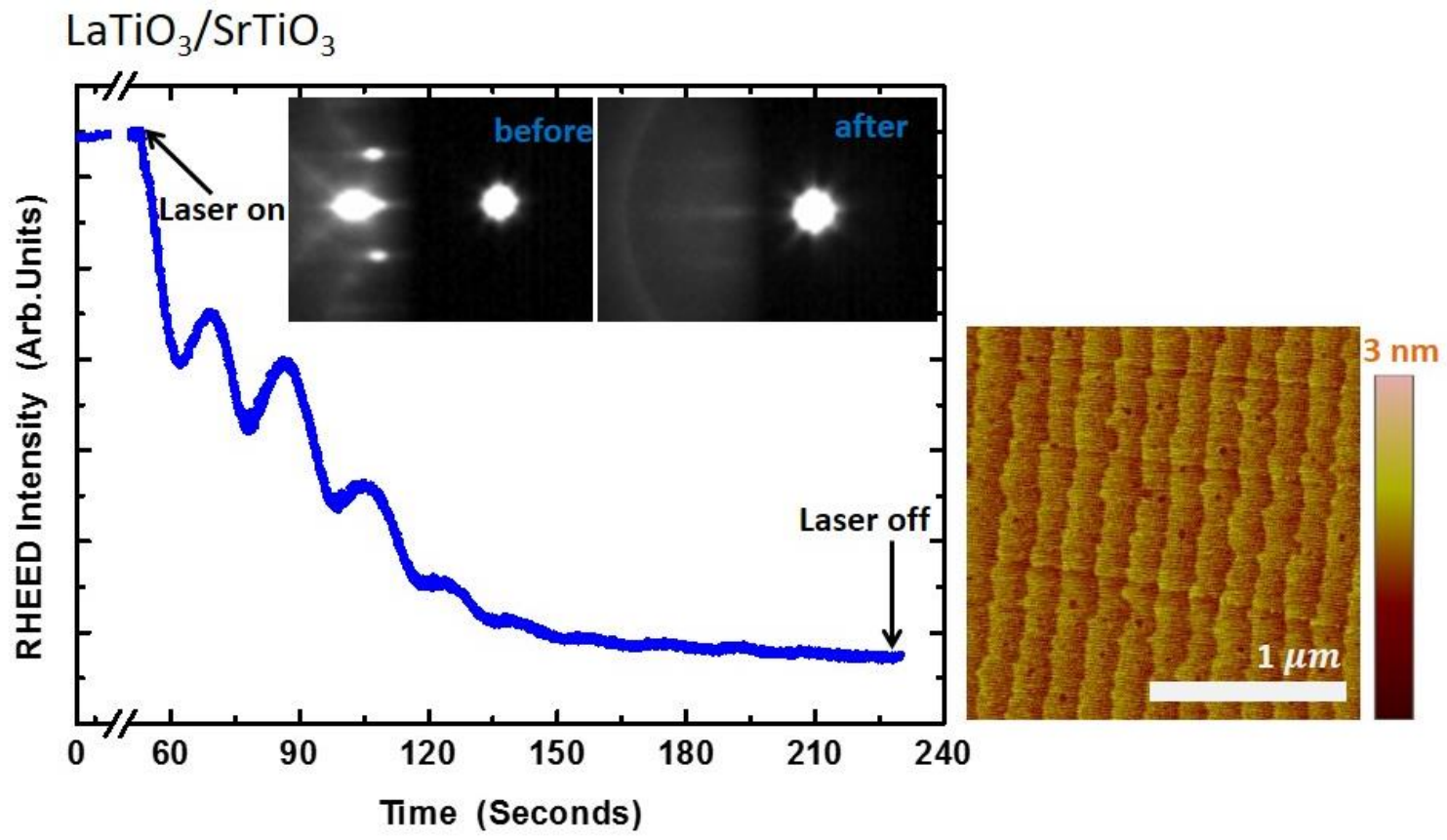

Figure S6. $LaTiO_3/SrTiO_3$

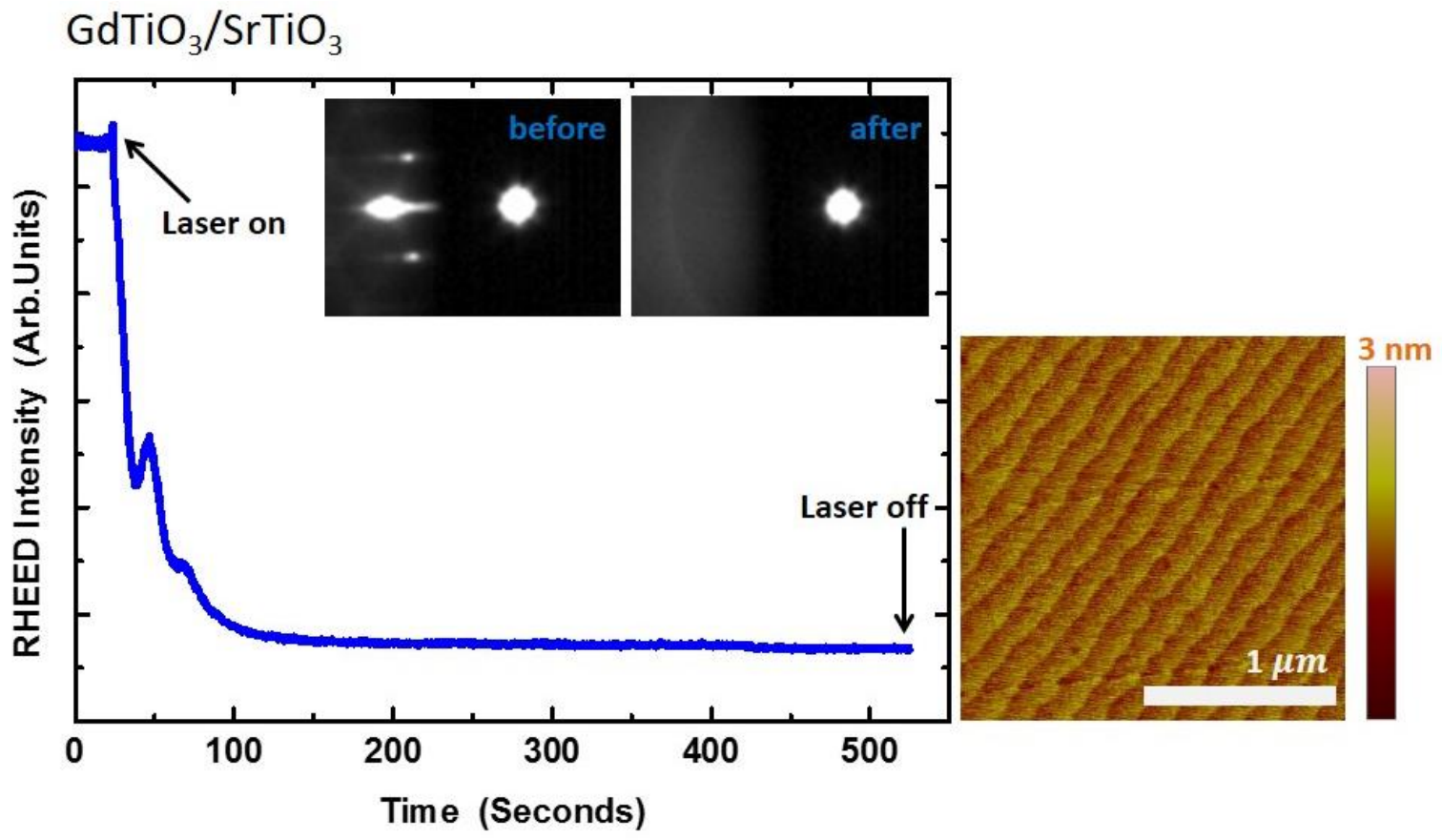

Figure S7. $GaTiO_3/SrTiO_3$

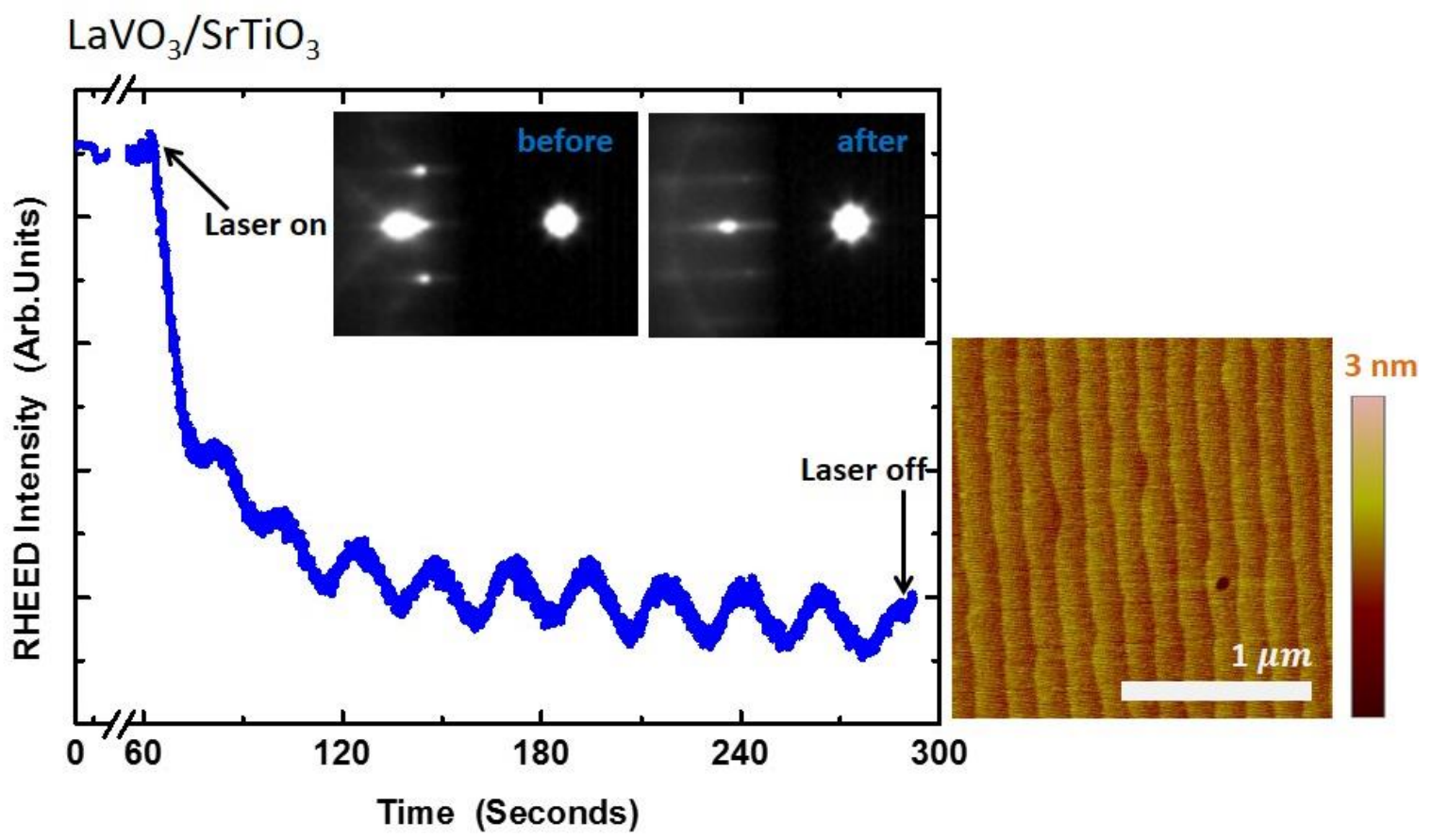

Figure S8. $LaVO_3/SrTiO_3$

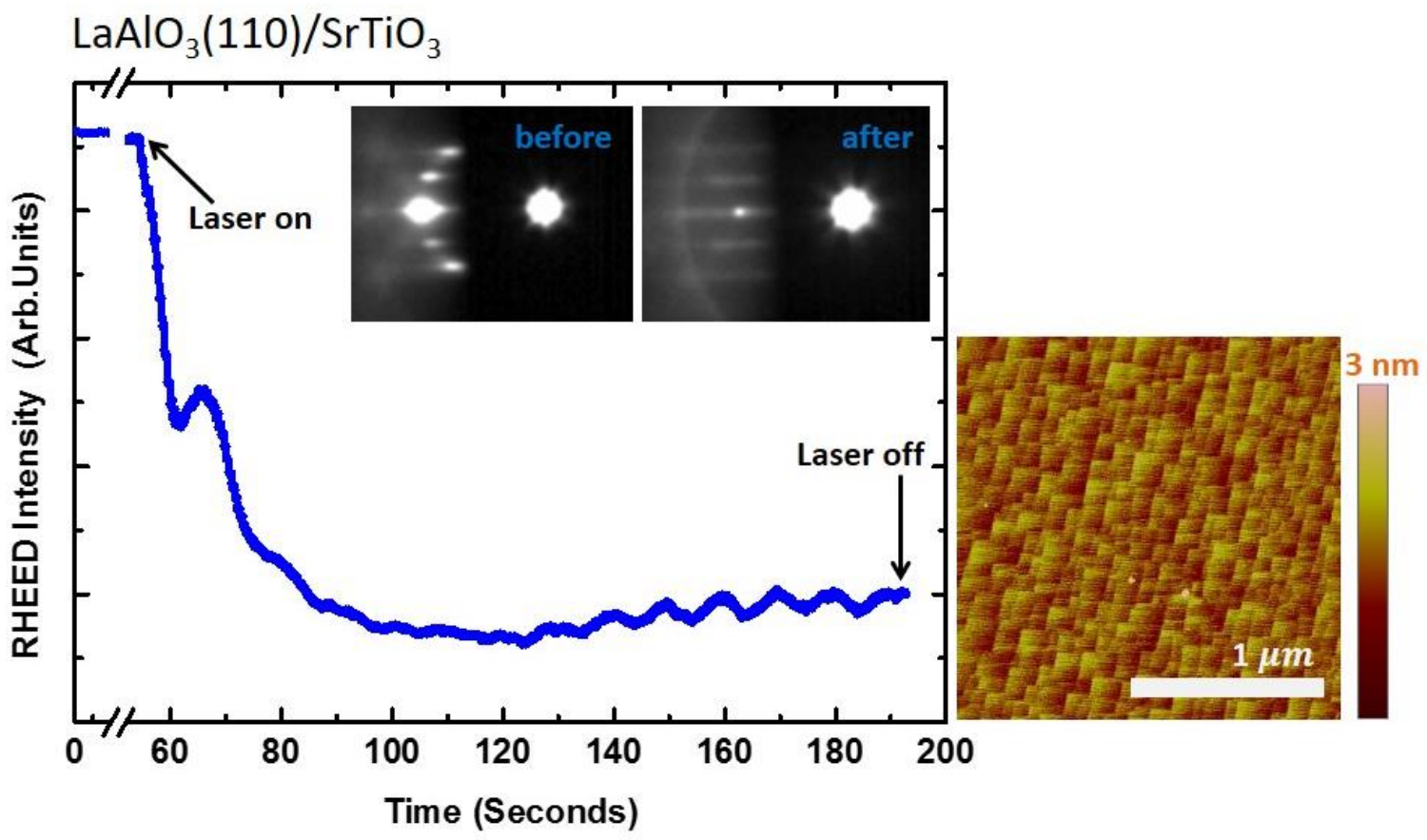

Figure S9. $LaAlO_3(110)/SrTiO_3$

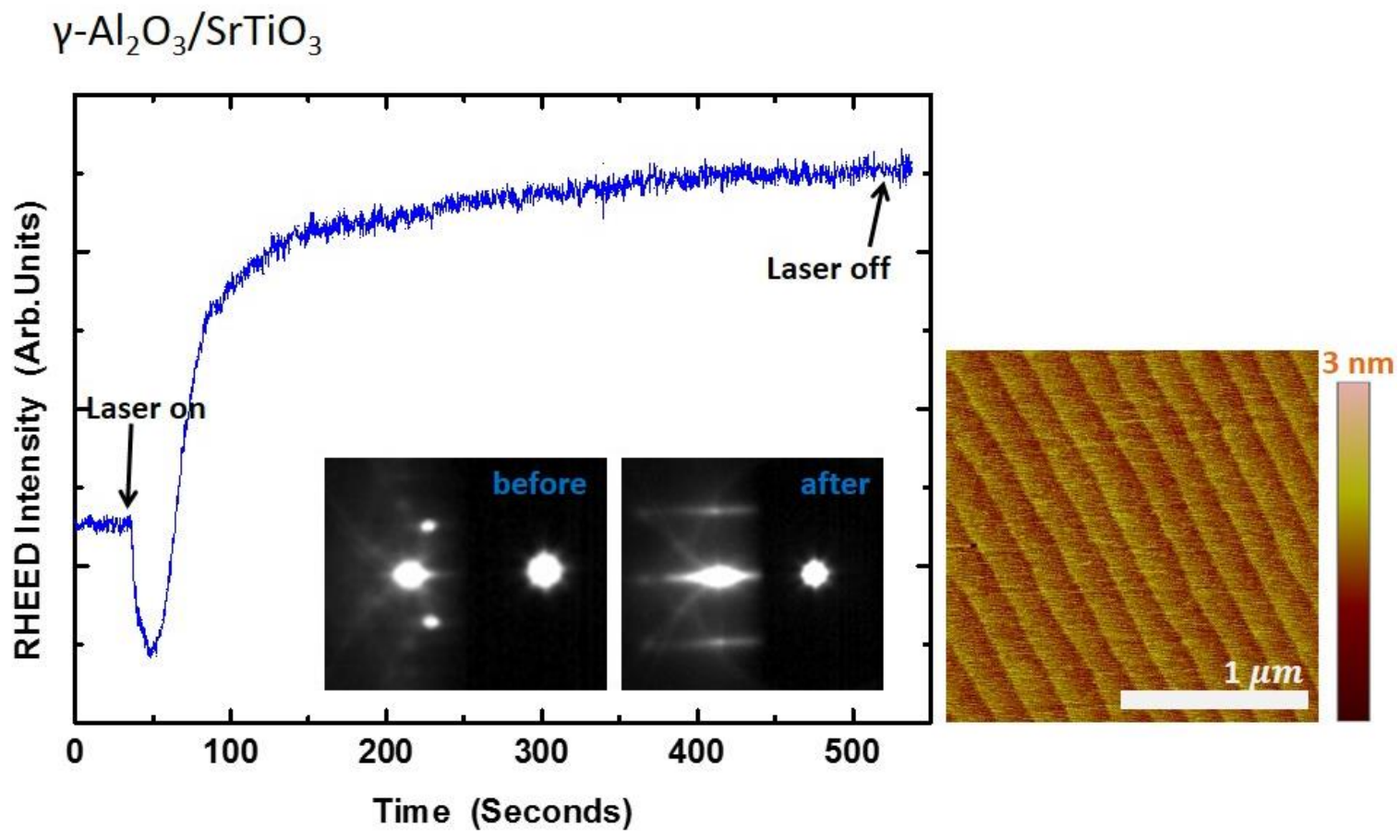

Figure S10. $\gamma$-$Al_2O_3/SrTiO_3$

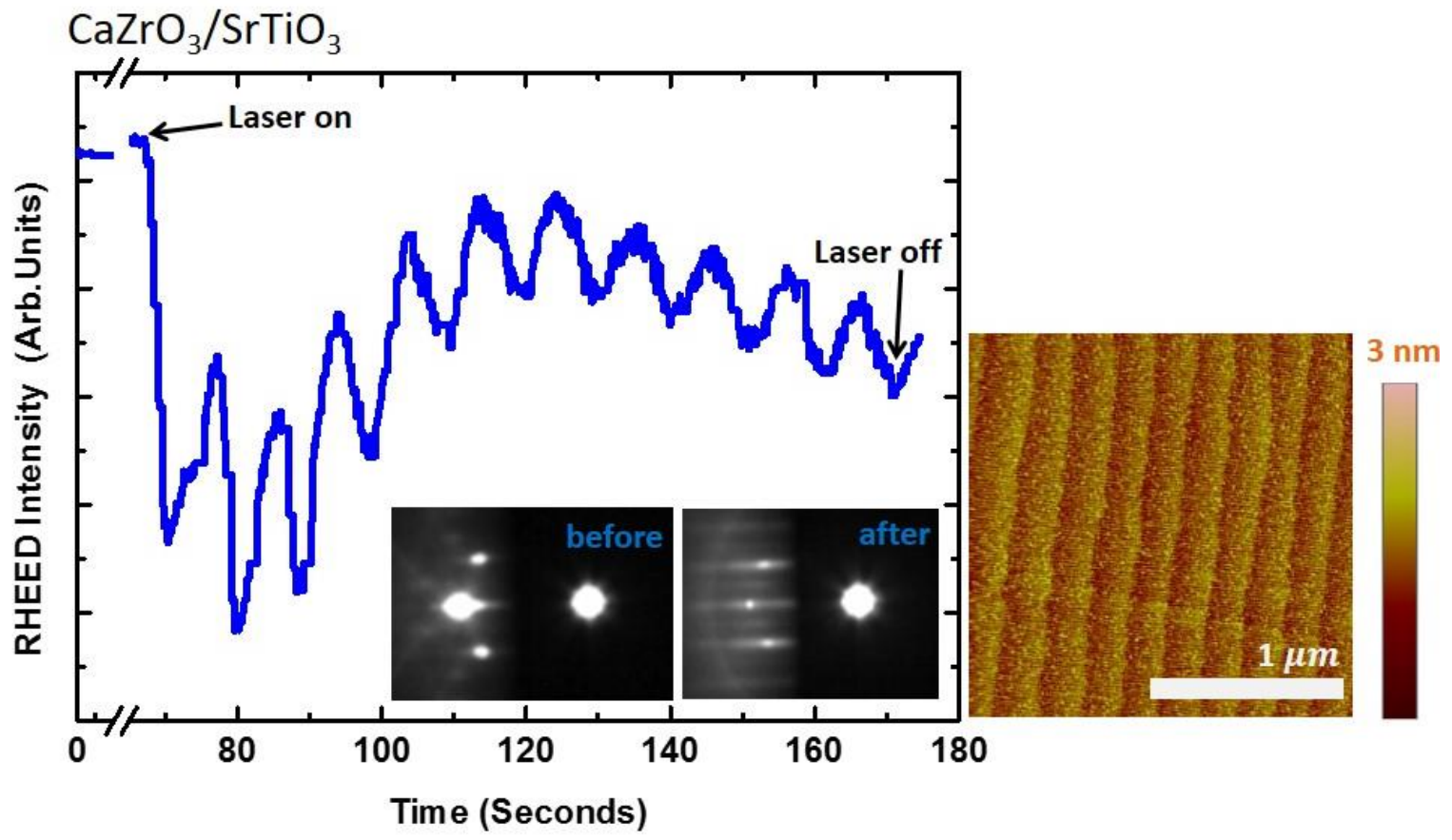

Figure S11. $CaZrO_3/SrTiO_3$

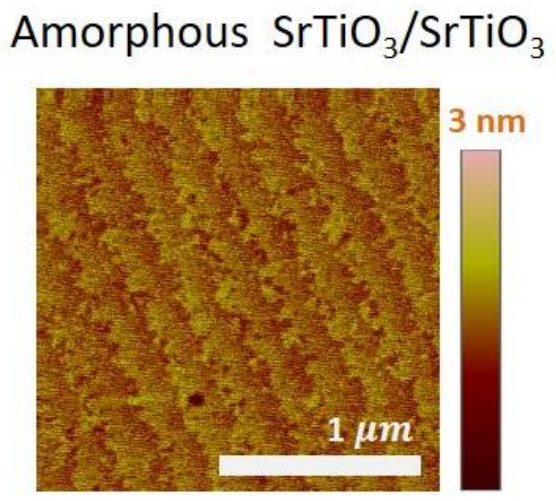

Figure S12. amorphous-$SrTiO_3/SrTiO_3$

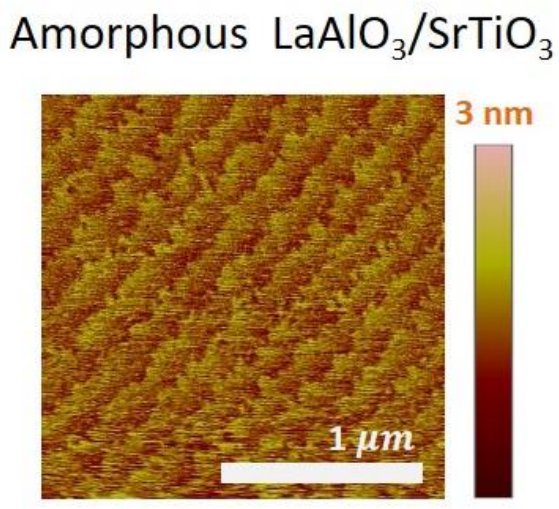

Figure S13. amorphous-$LaAlO_3/SrTiO_3$

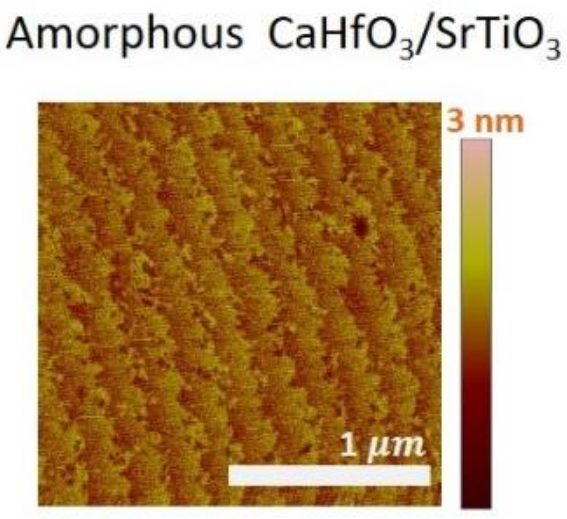

Figure S14. amorphous-$CaHfO_3/SrTiO_3$

**2. Surface treatments.**

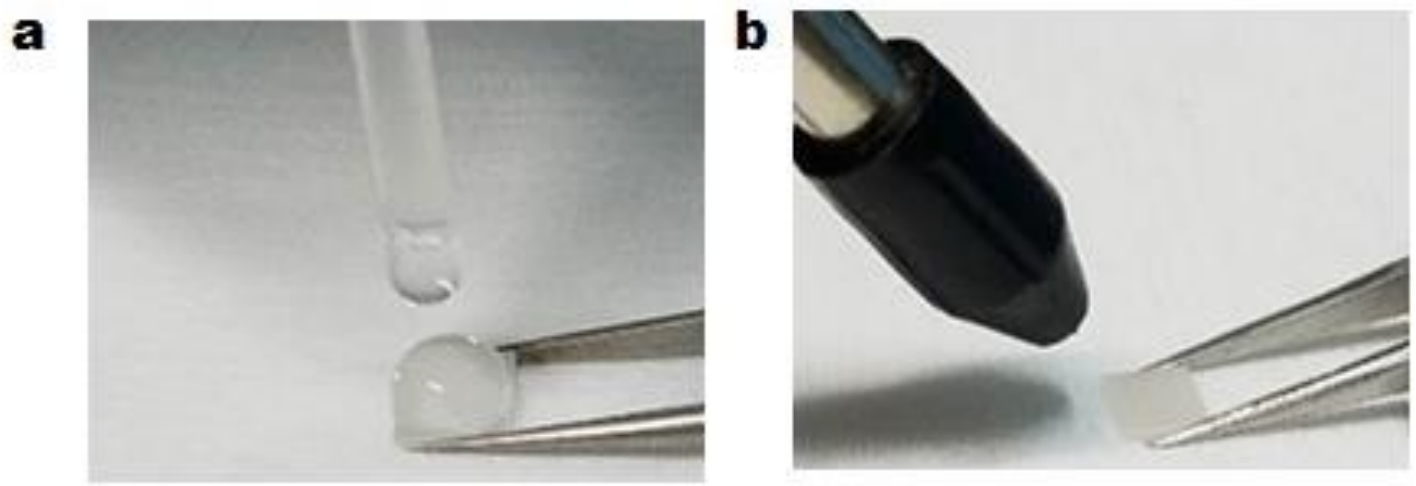

Figure S15. Pictures of surface treating processes. Liquid solvents were (a) dropped on the samples with a pipette, and then (b) dried immediately by blowing with a nitrogen gas gun.

It is known that, like water, ethanol [1] and acetone [2] can also be chemically adsorbed on oxides. The interaction between surface oxygen vacancies and ethanol (or acetone) may also involve chemistry that can provide SSL. Therefore, similar tuning effects are expected by surface treatments with water, ethanol, and acetone. A set of typical results are shown in Figure S16.

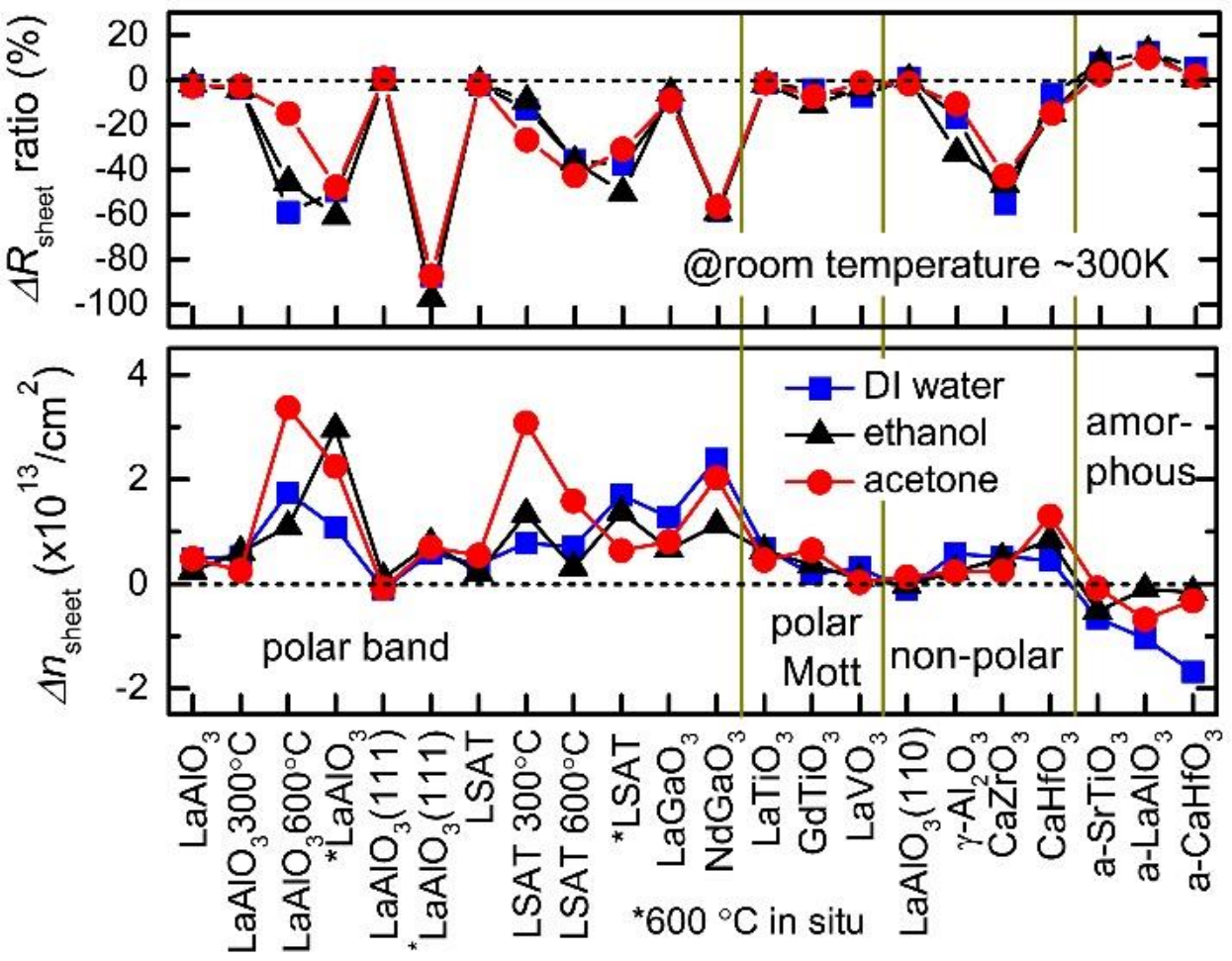

Figure S16.

### 3. An explanation for the distinct BE shift.

As schemed in Fig. 5B in the main text, the measured BE in AP-XPS is defined as the energy difference from core levels to the Fermi level, $E_F$. The core levels will follow the same tilting as that in conduction and valence bands. A simple electrostatic analysis shows that the transferred electrons will tilt the bands in $LaAlO_3$ downwards, but have no effect on the bands in $SrTiO_3$ (if ignoring the spatial distribution of the transferred electrons in $SrTiO_3$ whose effect is a band bending in $SrTiO_3$). Taking into account the spatial distribution of the transferred electrons, the bands in $SrTiO_3$ will be bended upwards (see Section 4 below). Due to the large dielectric constant of $SrTiO_3$, the band bending is small [3] compared with the potential change in $LaAlO_3$ induced by electron transfer. This explains why the BE shifts of core levels in $SrTiO_3$ and in $LaAlO_3$ are distinctly different (see Fig. 4D in the main text), which in turn is a signature of electron transfer from surface to interface.

### 4. More considerations of band alignment.

The valence band maximum (VBM) is roughly aligned at the interface since the valence bands of both $SrTiO_3$ and the capping oxides are derived from O 2p orbitals (for Mott insulators, refer to the band derived from O 2p orbitals rather than the low Hubbard band). The conduction band minimum (CBM) of the capping oxides is higher at the interface since its band gap (for Mott insulators, refer to the charge transfer gap [4]) is generally much wider than that of $SrTiO_3$.

In the $SrTiO_3$ side, depending on the detailed growth and annealing conditions, there might be some oxygen vacancies inside, more or less (red dots in Fig. S18a). From an electrostatic consideration the electrons in the surface shallow levels (SSL) can transfer into the interface once $E_{drive} > 0$, until $E_{drive}=0$ or exhausting all the available SSL. The transferred electrons produce an electric field across the capping layer which tilts downwards the whole bands, including core levels (Fig. S18b). The electrons transferred to the interface will distribute in a thin $SrTiO_3$ layer close to the interface, which bends up the $SrTiO_3$ bands, including the levels of oxygen vacancies, and thus creates a potential well near the interface. Consequently, some trapped electrons in oxygen vacancies in $SrTiO_3$ whose levels have been lifted higher than $E_F$ will move into the interfacial potential well, which in turn tunes down the $SrTiO_3$ bands until finally a balance is reached. The interfacial conductivity should be determined by the interplay of all the above processes. In addition, localizations of the transferred electrons may affect the interfacial conductivity as well, but will not obviously influence the charge transfer.

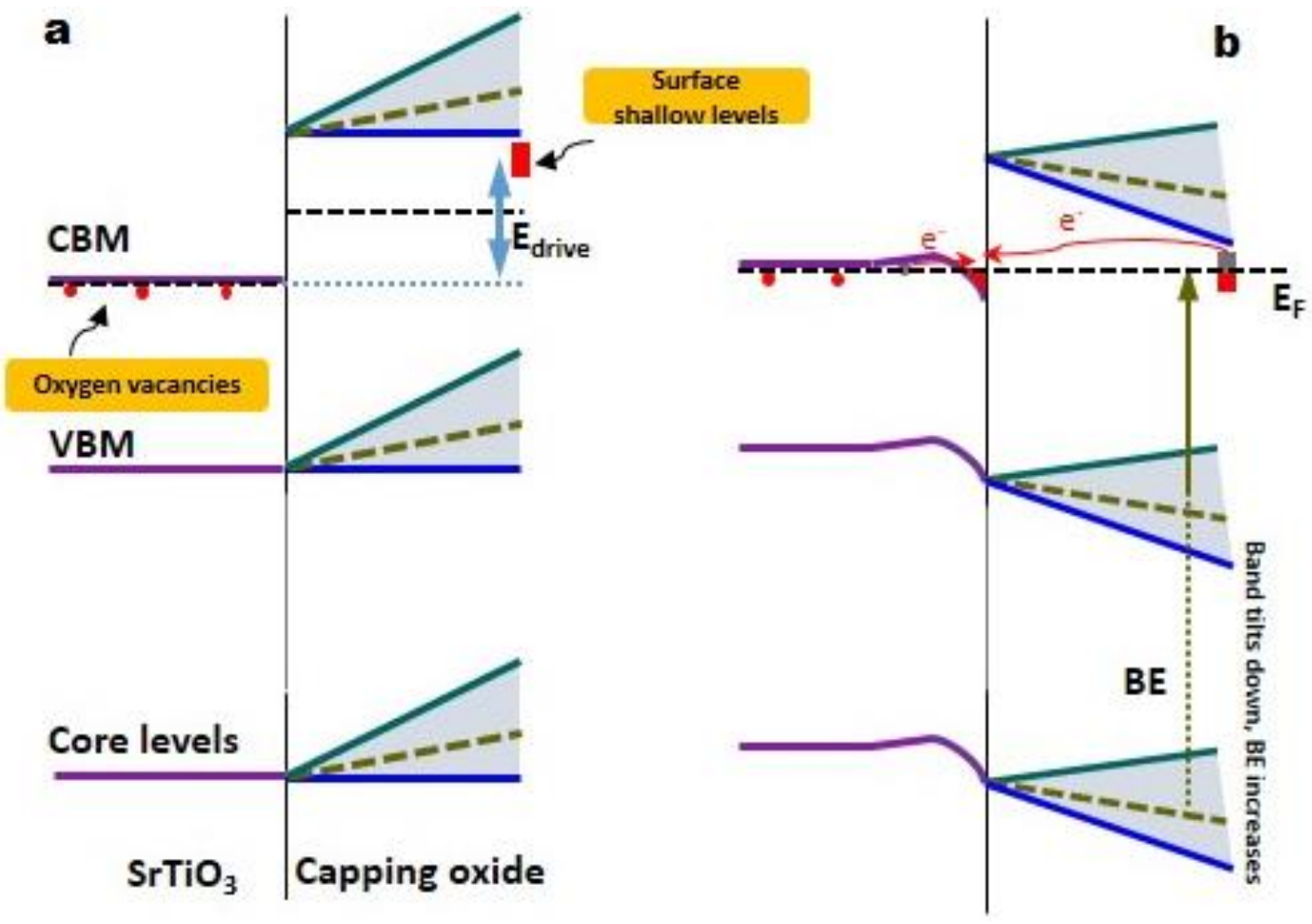

Figure S17. Schematic picture of charge transfer and band alignment. (a) Before and (b) after transferring electrons. Note that the $E_F$ in (a) is not balanced since it is only an imaginary situation before charge transfer. The band in capping layer is either flat or tilting up, depending on the polarity condition of the capping layer. The transferred charges, either from the surface or from the bulk, will generate an electrical field which will change the corresponding bands, accordingly. $E_{drive}$ is not strictly defined because SSL will distribute in an energy region rather than a single energy level. Binding energy (BE) is the energy difference between a core level and $E_F$.

In the present discussion, we didn't include the possible shallow levels in the bulk of capping oxide layers since they are less relevant (except for the oxygen vacancies in amorphous oxides, which may provide electrons to the interface). We also did not discuss the possibility that the surface oxygen vacancies themselves work as SSL. Interface mixture and structure defects were also not discussed. Although contributions from all these possibilities cannot be excluded, we emphasize that they only affect the magnitude of interfacial conductivity, and can coexist with the present scenario.

**5. Estimation of $E_{\mathrm{drive}}$.**

As defined in the main text, the $E_{\mathrm{drive}}$ must be related to the details of surface chemistry. Unfortunately, at present stage there are no such information available. Inspired by the fact that the amorphous-$SrTiO_3$/$SrTiO_3$ is conducting but its room-temperature $R_{\mathrm{sheet}}$ is quite high (equivalently $n_{\mathrm{sheet}}$ is low) (Fig. 1a in the main text), we speculated that in this case SSL are slightly above CBM of $SrTiO_3$ (so the initial $E_{\mathrm{drive}}$ is larger than, but not too far from 0). Except for γ-$Al_2O_3$, all the capping materials have the same perovskite structure. So we expect that the same speculation can be extended, and the energy difference between the CBM at the surface of capping oxide and that of $SrTiO_3$ substrate can be regarded as a crude estimation of the initial $E_{\mathrm{drive}}$ (before electron transfer). In this way, if the capping layer is non-polar, $E_{\mathrm{drive}}$ is the difference in CBM of $SrTiO_3$ and capping oxides, which is either directly cited from references [5–9], if available, or calculated from the reported band gap of capping oxides [10–12], assuming that the O 2p bands are aligned and the band gap of $SrTiO_3$ is 3.2 eV. If the capping layer is polar, additional contribution should be added. A built-in potential of ~1 eV has been detected in an insulating $LaCrO_3$/$SrTiO_3$ heterointerface where no surface-interface electron transfer has occurred [13]. We arbitrarily added 1 eV to all heterointerfaces with polar capping layers. The estimated $E_{\mathrm{drive}}$ is as shown in Figure 3b in the main text.